\documentclass[twocolumn]{aastex63}

\usepackage{natbib}
\usepackage{CJK}
\usepackage{amsmath}
\usepackage{natbib}
\usepackage{graphicx}
\usepackage{amsmath}
\usepackage{booktabs}
\usepackage{longtable}
\usepackage{xspace}
\usepackage{hyperref}
\usepackage[T1]{fontenc}
\usepackage{lipsum}
\usepackage{comment}
\usepackage{xcolor}
\usepackage{chngcntr}
\usepackage{url}
\usepackage[utf8]{inputenc}
\usepackage{multirow}

\usepackage[normalem]{ulem}
\renewcommand{\added}[1]{\textcolor{blue}{#1}}
\renewcommand{\deleted}[1]{\textcolor{red}{\sout{#1}}}
\renewcommand{\replaced}[2]{\textcolor{blue}{#1}{\textcolor{red}{\sout{#2}}}}

\newif\iftrackchanges
\trackchangesfalse 

\iftrackchanges
    \renewcommand{\added}[1]{\textcolor{blue}{#1}}
    \renewcommand{\deleted}[1]{\textcolor{red}{\sout{#1}}}
    \renewcommand{\replaced}[2]{\textcolor{blue}{#1}{\textcolor{red}{\sout{#2}}}}
\else
  \renewcommand{\added}[1]{#1}
  \renewcommand{\deleted}[1]{}
  \renewcommand{\replaced}[2]{#1}
\fi

\received{}
\revised{}
\accepted{}

\shorttitle{Debris disk in CPD -72 2713}
\shortauthors{Zhang et al.}

\graphicspath{{./}{figures/}}

\begin{document}
\begin{CJK*}{UTF8}{gbsn}

\title{A Misaligned Debris Disk relative to the Host Star CPD-72 2713 from JWST/NIRCam Imaging}


\correspondingauthor{Jingwen Zhang}
\email{jwzhang@ucsb.edu}

\author[0000-0002-2696-2406]{Jingwen Zhang (张婧雯)}
\affiliation{Department of Physics, University of California, Santa Barbara, CA 93106, USA}

\author[0000-0001-6205-9233]{Maxwell A. Millar-Blanchaer}\affiliation{Department of Physics, University of California, Santa Barbara, CA 93106, USA}

\author[0000-0002-6964-8732]{Kellen Lawson}
\affiliation{Center for Space Sciences and Technology, University of Maryland, Baltimore County, Baltimore, MD 21250, USA}

\affiliation{Astrophysics Science Division, NASA-GSFC, Greenbelt, MD 20771, USA}
\affiliation{Center for Research and Exploration in Space Science and Technology, NASA-GSFC, Greenbelt, MD 20771, USA}

\author[0000-0002-0834-6140]{Jarron Leisenring}
\affiliation{Department of Astronomy/Steward Observatory, University of Arizona, 933 N. Cherry Ave., Tucson, AZ 85721, USA}

\author[0000-0003-2649-2288]{Brendan P. Bowler}
\affiliation{Department of Physics, University of California, Santa Barbara, Santa Barbara, CA 93106, USA}

\author[0000-0001-5365-4815]{Aarynn L. Carter}
\affiliation{Space Telescope Science Institute (STScI), 3700 San Martin Drive, Baltimore, MD 21218, USA}

\author[0000-0003-4614-7035]{Beth Biller}\affiliation{Institute for Astronomy, University of Edinburgh, Royal Observatory, Edinburgh EH9 3HJ, UK}
\affiliation{Centre for Exoplanet Science, University of Edinburgh, Edinburgh EH8 9YL, UK}

\author[0000-0001-5653-5635]{Tim D. Pearce}\affiliation{School of Physics, Trinity College Dublin, The University of Dublin, College Green, Dublin 2, Ireland}
\affiliation{Department of Physics, University of Warwick, Gibbet Hill Road, Coventry CV4 7AL, UK}

\author[0000-0002-5352-2924]{Sebastian Marino}\affiliation{Department of Physics and Astronomy, University of Exeter, Stocker Road, Exeter EX4 4QL, UK}

\author[0000-0002-7325-5990,gname='Ellis',sname='Bogat']{Ellis Bogat}
\affiliation{Department of Astronomy, University of Maryland, College Park, MD 20782, USA}
\affiliation{NASA Goddard Space Flight Center, 8800 Greenbelt Road, Greenbelt, MD 20771, USA}
\affiliation{Center for Research and Exploration in Space Science and Technology, NASA Goddard Space Flight Center, Greenbelt, MD 20771, USA}

\author[0000-0002-4918-0247]{Robert J. De Rosa}
\affiliation{European Southern Observatory, Alonso de C\'{o}rdova 3107,
Vitacura, Casilla 19001, Santiago, Chile}

\author[0000-0002-9173-0740]{\'{E}lodie Choquet}
\affiliation{Aix Marseille Universit\'{e}, CNRS, CNES, LAM, Marseille, France.}

\author[0000-0001-8627-0404]{Julien H. Girard}
\affiliation{Space Telescope Science Institute (STScI), 3700 San Martin Drive, Baltimore, MD 21218, USA}

\author[0000-0003-0958-2150]{Nicolas Godoy}
\affiliation{Aix Marseille Universit\'{e}, CNRS, CNES, LAM, Marseille, France.}

\author[0000-0001-8074-2562]{S.~Hinkley}
\affiliation{University of Exeter, Physics Building, Stocker Road, Exeter EX4 4QL, United Kingdom}

\author[0000-0003-1212-7538]{Bruce Macintosh}
\affiliation{Department of Astronomy and Astrophysics, UC Santa Cruz, Santa Cruz CA 95064}

\author[0000-0003-0593-1560]{Elisabeth C. Matthews}\affiliation{Max-Planck-Institut f\"{u}r Astronomie, D-69117 Heidelberg, Germany}

\author[0000-0003-2259-3911]{Shrishmoy Ray}\affiliation{School of Mathematical \& Physical Sciences, Macquarie University, 12 Wally’s Walk, Macquarie Park, NSW 2113, Australia}

\author[0009-0007-6432-0328]{Connor Vancil}\affiliation{Department of Physics, University of California, Santa Barbara, CA 93106, USA}

\author[0009-0004-5080-2906]{Lisa Altinier}
\affiliation{Aix Marseille Universit\'{e}, CNRS, CNES, LAM, Marseille, France.}



\begin{abstract}

Observations of debris disks provide a valuable window to investigate the formation and evolution of planetary systems.
We present high-contrast \textit{JWST}/NIRCam observations of CPD-72 2713, a 24 Myr-old  star that belongs to the $\beta$ Pic moving group. We obtained the first scattered-light images of the disk in both F210M and F444W filters, revealing a nearly face-on orientation. Forward modeling yields disk radii of  $\rm 55.4^{+3.0}_{-3.2}\,AU$ (F210M) and  $\rm 39.1^{+4.6}_ {-2.6}\,AU$ (F444W), both significantly smaller than the thermal emission radii measured with ALMA and Herschel/PACS. This discrepancy suggests that scattered light traces  the inner region of the disk while thermal emission probes the outer regions. Combining the stellar rotation period ($4.47\pm0.67$ days) from TESS light curves and spectroscopic $v\sin i_*$, we constrained the stellar inclination to  $47^{\circ}.2^{+11.1}_{-7.7}$ ($1\sigma$). The minimum star-disk mutual inclinations are  $27^{\circ}.6 _{-8.2} ^{+11.5}\ (1\sigma)$ from F210M data and  $27^{\circ}.3 _{-10.9}^{ +12.5}\ (1\sigma)$ in F444W data. We find that the minimum star--disk mutual inclination exceeds $10^\circ$ at the $95\%$ credible level in both F210M and F444W, suggesting a misalignment between the scattered-light disk and the stellar equatorial plane.  We note that variation among published disk inclination measurements and potential systematic uncertainties may affect this probability. With JWST high-contrast imaging, we rule out planets down to $0.3\ \rm M_{J}$ at $0.5^{\prime\prime}$ and beyond, disfavoring a planet as the driver of the star-disk misalignment.


\end{abstract}


\section{Introduction}\label{sec:intro}
Debris disks are circumstellar belts composed of dust, small rocky or icy bodies, and planetesimals-remnants of the planet formation process. Unlike gas-rich protoplanetary disks around very young stars ($\lesssim 10$ Myr), debris disks are generally gas-poor and optically thin, which makes them more effective tracers of disk substructures shaped by planets \citep{Hughes2018, Pearce2024}. High-resolution imaging has revealed a variety of such features, including rings, gaps \citep{Marino2018}, warps \citep{Hales2022}, and clumps \citep{Booth2023}, many of which can be explained by secular or resonant interactions with unseen planets \citep{Ida2000,Wyatt2003,Mustill2012,Pearce2014}. A classic example is the warp in the $\beta$ Pictoris disk, initially predicted to be caused by a planet on an inclined orbit \citep{Heap2000} and later confirmed with the discovery of $\beta$ Pictoris b \citep{Lagrange2010}.

Planets can also sculpt debris disks dynamically by scattering or accreting material, producing sharp edges rather than diffuse boundaries \citep{2019AJ....158..162F, ImazBlanco2023, Pearce2024}. In our own Solar System, Neptune exerts a similar role, dominating the orbital evolution of Kuiper Belt objects and shaping both the inner and outer edges of the belt \citep{Levison2008}. Extending this idea, \cite{Pearce2022} inferred the presence and properties of hypothetical planets in 178 debris disk systems by modeling the locations and shapes of their inner edges.

Another key property of debris disks is that their dust must be continually replenished through destructive collisions of larger planetesimals. This requires a stirring mechanism to excite planetesimal orbits. While stirring by the disc's self gravity is possible, it can sometimes require an unphysically massive disc \citep{Krivov2018,Krivov2021,Pearce2022}. Alternatively,  planet-disk interactions can also provide the necessary stirring, exciting random velocities of debris particles to trigger collisions, \added{known as "planetary stirring" \citep{Mustill2009}}. In this context, debris disks are not only relics of planet formation but also powerful signposts of hidden planets.

 While they can persist throughout a star's lifetime, debris disks are most commonly resolved around young stars, where they are brighter and easier to detect \citep{Hughe2018}. In addition many debris disks lie in the outer regions of planetary systems, at tens to hundreds of AU. These features  coincide with the sensitivity range of high-contrast direct imaging, which is optimized for detecting young, wide-orbit planets. Systems where both planets and debris disks are detected offer valuable opportunities to study their formation and interactions. However, the number of systems in which both planets and debris disks have been directly imaged remains small \citep{Hughe2018}, reflecting the challenges of directly imaging both planets and disks.  Previous direct imaging surveys suggest a potentially higher occurrence rate of $5-20\ \rm M_{Jup}$ planets at 10-1000 AU in systems with debris disks compared to those without ($6.27\%$ vs. $0.73\%$, \citealt{Meshkat2017}); however, this conclusion is based on a relatively small sample (seven planets in four debris disk systems).

Advances in instrumentation are pushing detection limits and increasing the likelihood of discovering such systems. The James Webb Space Telescope (JWST) currently offers the highest sensitivity for directly imaging planetary companions, capable of detecting planets with sub-Jupiter masses. \added{Using the coronagraphic mode with the Mid-Infrared Instrument (MIRI) on JWST, \citet{Lagrange2025} reported the discovery of a point source at the projected separation of $1.5^{\prime\prime}$ from TWA 7, which would provisionally place it inside the surrounding young debris disk if confirmed.} The source is consistent with a sub-Jovian planet with a mass of approximately $0.3\ \mathrm{M_{J}}$. A follow-up detection at the same location was later reported using JWST's Near-Infrared Camera (NIRCam), producing photometry consistent with a Saturn-mass planet in the shorter wavelength bandpasses (F200W and F444W, \citealt{Crotts2025}).

In this paper, we present NIRCam observations of the debris disk system, CPD -72 2713, in the F210M and F444W filters. CPD -72 2713 is a young, late-type star ($\sim 24$ Myr, K7v) in the $\beta$ Pic Moving Group, at a distance of 36.72 pc (see stellar parameters in Table~\ref{tab:star-compare}). The disk has been spatially resolved using Atacama Large Millimeter/submillimeter Array (ALMA, \citealt{Wootten2009}) and Photodetector Array Camera and Spectrometer (PACS, \citealt{Poglitsch2010}) on Herschel Space Observatory. These observations reveal that CPD -72 2713 has a large, wide, face-on and dust-rich disk \citep{Moor2020,Tanner2020,Marino2026} . In Section~\ref{sec:obs}, we described our NIRCam observations and data reduction methods. Section~\ref{sec:disks} detailed the forward modeling of the debris disks. We present the measurement of the star-disk mutual inclination in Section~\ref{sec:sd_misalign} and discuss possible mechanisms responsible for the misalignment in Section~\ref{sec:diss}. In Section~\ref{sec:plantes}, we describe our planet-search analysis and present the upper limits derived from the JWST contrast curves. We summarize our conclusions in Section~\ref{sec:conc}.

\begin{deluxetable}{lcc}
\tablecaption{Stellar Parameters of CPD 72-2713 (other names: TYC 9340-437-1, 2MASS J22424896-7142211, Gaia DR3 6382640367603744128 )}
\label{tab:star-compare}
\tablehead{\colhead{Parameters }&\colhead{Value}  &\colhead{Reference}  }
\startdata                        
$T_{\rm eff}\ (\rm{K})$  & $4091\pm{12}$    & a   \\
$M_{\star}$ ($\rm M_{\odot}$) & $0.753\pm{0.009}$ & a  \\
$R_{\star}$ ($\rm R_{\odot}$) &  $0.855\pm{0.006}$& a \\
Spectral type & K7IVe & b \\
Age (Myr)& $24\pm{3}$  & c   \\
$\varpi$ (mas)  & $27.23\pm{0.01}$  &  d  \\
$d$ (pc)  & $36.72\pm{0.01}$  &  d  \\
$L_{\rm dust}/L_\star$ & $1.04^{+0.03}_{-0.03}\times10^{-3}$& e \\
$v\sin i_*$ ($\rm km\ s^{-1}$) & $7.3\pm{1.0}$ &  f  \\
                               & $6.6\pm{1.4}$ &  g \\
                               & $7.5\pm{1.2}$ & h \\
                                & $5.9\pm{0.4}$ & j \\
$\mathrm{P_{rot}}$ (days)   & $4.47\pm{0.67}$ & This work    \\
                             & 4.437 & i   \\
$i_*$ (deg)&  $ 47.2^{+11.1}_{-7.7}$ &  This work   \\
&   $42\pm{7}$ &  i    \\
\enddata
\tablecomments{ a. \citet{Marino2026}; b. \cite{Pecaut2013}; c.\cite{Bell2015}; d. \cite{Gaia}; e. \cite{Matra2025}; f.\cite{Zakhozhay2022}; g. \cite{Weise2010}; h. \cite{Torres2006}; i.\cite{Moor2020}; j.\cite{ZF2021} } 
\end{deluxetable}


\begin{deluxetable*}{ccccccccccc}
\tablecaption{Summary of Target and Reference Exposure Settings}\label{tab:obs}
\tablehead{
\colhead{Name} & \colhead{Type} & \colhead{Filter} & \colhead{Mask} & \colhead{Readout} & 
\colhead{$\mathrm{N_{groups}}$} & \colhead{$\mathrm{N_{ints}}$} & 
\colhead{$\mathrm{t_{int} (s)}$} & \colhead{$\mathrm{t_{exp}  (s)}$} &
\colhead{$\mathrm{N_{dithers}}$} & \colhead{$\mathrm{N_{rolls}}$}
}
\startdata
CPD 72-2713 & SCI & F210M & MASK335R & DEEP8    & 7 & 5 & 136.8     & 684.2     & 1 & 2 \\
CPD 72-2713 & SCI & F444W & MASK335R & DEEP8    & 7 & 5 & 136.8     & 684.2     & 1 & 2 \\
HD 210742   & REF & F210M & MASK335R & SHALLOW4 & 7 & 6 &  36.3  & 218.1   & 5 & 1 \\
HD 210742   & REF & F444W & MASK335R & SHALLOW4 & 7 & 6 & 36.3  & 218.1   & 5 & 1 \\
\enddata
\tablecomments{$\mathrm{N_{ints}}$ is the total number of integrations, 
$\mathrm{N_{groups}}$ is the number of groups per integration, 
and $\mathrm{N_{dithers}}$ is the number of dithers, 
$\mathrm{t_{int}}$ is the effective integration time, 
$\mathrm{t_{exp}}$ is the total exposure time.} 
\end{deluxetable*}

\section{Observations $\&$ Data Reduction} \label{sec:obs}

\subsection{Observations}

The observation of CPD -72 2713 \footnote{The JWST data presented in this article were obtained from the Mikulski Archive for Space Telescopes (MAST) at the Space Telescope Science Institute. The specific observations analyzed can be accessed via \dataset[doi: 10.17909/e4s2-5t93]{https://doi.org/10.17909/e4s2-5t93}.} was part of program "Finding the great sculptors: A Renaissance in Planet Disk
Dynamics" (GO 6102, PI: Millar-Blanchaer), with the goal of searching for debris-sculpting planets. The observation was executed on 2024-07-23 UT with the JWST/NIRCam coronagraphic imaging mode \citep{Girard2022}. The target was observed in dual-channel mode using the F210M ($\rm \lambda_{ref}=2.10\ \mu m$, $\Delta \lambda = 0.27\ \mu m$) and F444W ($\rm \lambda_{ref}=4.40\ \mu m$, $\Delta \lambda = 1.02\ \mu m$) filters with the intermediate round occulter MASK335R. The F210M data were obtained with a field of view (FOV) of $10^{\prime\prime} \times 10^{\prime\prime}$ and a pixel scale of $0.031^{\prime\prime}$ per pixel, while the F444W data covered a FOV of $20^{\prime\prime} \times 20^{\prime\prime}$ with a pixel scale of $0.063^{\prime\prime}$ per pixel. These two filters were specifically chosen to help identify background contaminants, as planets are significantly brighter at 4 $\micron$ than at 2 $\micron$, whereas Galactic O to M stars exhibit much smaller differences in brightness at the two wavelengths \citep{Skemer2014}.

All observations were repeated at two
independent roll angles separated by $\sim 9^{\circ}$ to enable subtraction of the residual stellar point-spread function (PSF) through angular differential imaging (ADI). In addition, we observed a bright single reference star, HD 210742, with identical instrument settings to enable reference differential imaging (RDI). The reference star was chosen for its similar spectral type to the science target and its proximity (within $10^{\circ}$), minimizing telescope slew overheads and reducing position-dependent wavefront drift between science and reference observations. Each reference observation was repeated at five distinct dither positions, following the 5-POINT-BOX small-grid pattern. This approach was designed to build a small library of reference PSFs that capture varying degrees of misalignment between the star and the coronagraphic mask center. Table~\ref{tab:obs} presents a summary of the observations used in this paper.

\subsection{Data Reduction}

We performed the data reduction using the open-source \texttt{Python} package \texttt{SpaceKLIP}\footnote{\url{https://spaceklip.readthedocs.io/en/latest/}} \citep{Kammerer2022, Carter2023}, which runs the official \texttt{jwst}\footnote{\url{https://jwst-pipeline.readthedocs.io/en/latest/}} Science Calibration Pipeline \citep{Bushouse2022} with parameter customizations tailored for high-contrast imaging. The reduction process generally consists of three stages. In Stage 1, the pipeline applies detector-level corrections to the raw, uncalibrated data, including saturation, superbias, non-linearity, and dark current, followed by ramp fitting. This produces calibrated count-rate images. We disable the $1/f$ noise correction at this stage, as it can introduce artifacts in NIRCam images, especially for disk targets. In Stage 2, we applied additional corrections and calibrations to the count-rate products from Stage 1 to generate fully calibrated outputs. This includes background subtraction, flat-fielding, and photometric calibration. 

In Stage 3, we applied median subtraction, bad-pixel correction, and frame re-centering and alignment. The bad-pixel correction involved several steps: (1) pixels flagged by the data quality (DQ) array (e.g., indicating hot or warm pixels, unreliable data processing) were replaced with the median of their orthogonal and diagonal neighbors; (2) bad pixels were identified and corrected based on temporal variations across integrations using a $6\sigma$ threshold; (3) additional bad pixels were flagged from spatial variations across integrations using a $5\sigma$ threshold; and (4) we provided a custom bad-pixel map generated through virtual checks, correcting flagged pixels using the same replacement method as for DQ-flagged pixels. Image centering and alignment begin by \replaced{determining the position of the star relative to the center of the coronagraphic mask in the first science frame.}{calculating the offset between the star's position and the center of the coronagraphic mask.} This is done by cross-correlating a simulated coronagraphic PSF generated with \texttt{stpsf}\footnote{\url{https://stpsf.readthedocs.io/en/latest/usage.html}} \citep{Perrin2014} against the observed PSF. \added{ Then the algorithm will calculate  the shift required to align all other frames with the first science frame and shift all frames to place the star at the geometric center of the array.} \deleted{The shift required to center the star is derived from the first science (SCI) frame. Subsequent shifts for the remaining science and reference images are then calculated relative to this first SCI frame using the same cross-correlation approach.}

\begin{figure*}
    \centering
    \includegraphics[width=\linewidth]{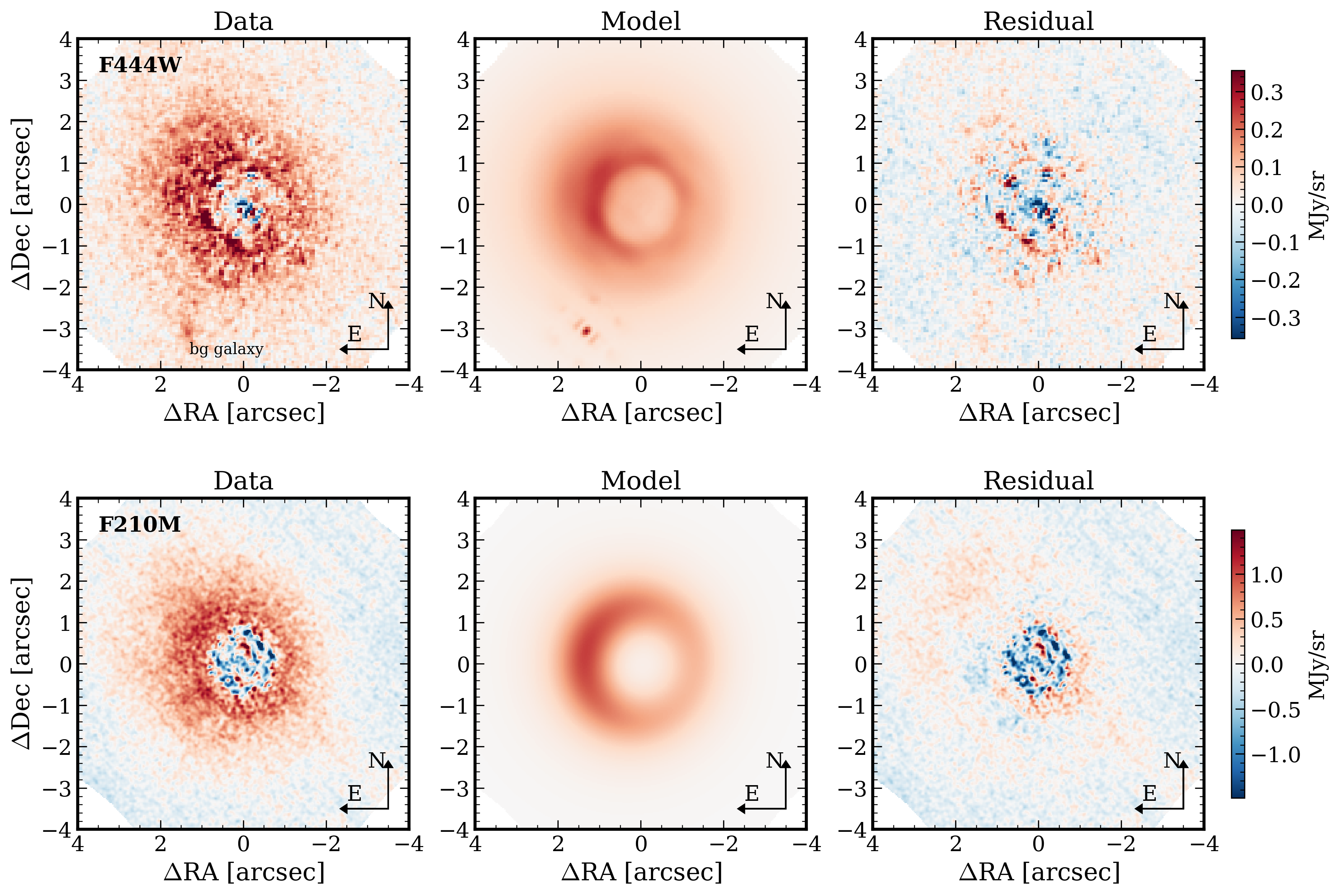}
    \caption{NIRCam observations and forward-modeling results for CPD -72 2713 using \texttt{Winnie} in the F444W (top) and F210M (bottom) filters. The left panels show MCRDI data reductions, the middle panels display the best-fitting forward models, and the right panels present the corresponding residual maps. In the F444W filter, we simultaneously modeled the disk and a source (background galaxy), whereas in the F210M filter, only the disk was modeled. Both images are rotated with North up and East left. The star is centered in the frame. Surface brightness is scaled linearly in both filters. }
    \label{fig:figure1}
\end{figure*}

\subsection{PSF Subtraction}
After completing the three reduction stages, the data are ready for PSF subtraction. First, we use \texttt{SpaceKLIP} to perform PSF subtraction in both filters using three principal component analysis (PCA) approaches: (1) ADI subtraction, using the two independent roll angles of CPD -72 2713; (2) RDI subtraction, using the observations of reference star HD 210742 to build the PSF library; and (3) combined ADI+RDI subtraction, which is identical to the RDI method but also includes opposite-roll images in the PSF library. In all cases, the subtraction is applied to each integration from both science rolls separately, followed by rotation to a common orientation (north-up and east-left) and co-addition. All subtractions are performed using a single annulus and a single azimuthal subsection. We used the number of KL modes as 1, 2, 5, 10, 20, \replaced{35}{50}. 

Next, to complement the \texttt{SpaceKLIP} reductions and explore potential differences in disk morphology and sensitivity, we also performed RDI subtraction on the reduced data using \texttt{Winnie}\footnote{ We used Winne version v1.1.0-3-gdecb4d4, corresponding to Git commit decb4d43fab1171cfdbd6ac836130ac3cc566ae2. The GitHub repository is available at \url{https://github.com/kdlawson/Winnie}. } \citep{Lawso2022,Lawson2023}, a Python package developed specifically for PSF subtraction and forward modeling of circumstellar disks in high-contrast JWST imaging. \texttt{Winnie} first calculates a set of coefficients for each reference image frame and constructs the stellar PSF model by multiplying each reference image by its corresponding coefficient and summing the results. There are three types of RDI methods available in \texttt{Winnie}. In standard RDI, the coefficients are chosen to minimize the squared residuals between the model PSF and the science image. In high-pass filtered RDI (HPFRDI), the coefficients are optimized using high-pass filtered versions of both the science and reference images. In model-constrained RDI (MCRDI), the coefficients are determined by minimizing residuals between the reference PSF model and the science image after subtracting a disk model, minimizing the loss of disk signal.

Figure~\ref{fig:figure1} presents the MCRDI PSF-subtracted images in the F444W and F210M filters, revealing a clearly visible face-on disk in both. We also identified a source (see Figure~\ref{fig:figure1}) at roughly $\Delta \rm{RA} \sim +1.3^{\prime\prime}$ and $\Delta \rm{Dec}\sim-3.0^{\prime\prime}$ in F444W images. The source is not seen in F210M filter. We confirmed it as a background galaxy and will discuss more about the source in Section~\ref{sec:plantes}.

\begin{deluxetable*}{lccccccc}
\tablecaption{List of best fitting model parameters for disk
using MCMC sampling}\label{tab:rvs}
\tablehead{
\colhead{} & \multicolumn{2}{c}{F210M} & \colhead{}  &\multicolumn{2}{c}{F444W} \\
\hline
\colhead{Parameter (Unit)} & \colhead{Prior} & \colhead{Credible Interval}& \colhead{Best-fit}   & \colhead{Prior} & \colhead{Credible Interval}& \colhead{Best-fit}
}
\startdata
$r_0$ (AU) & $\mathcal{U}(10, 130)$ & $55.4^{+3.0}_{-3.2}$ & 54.3  & $\mathcal{U}(10, 130)$ & $39.1^{+4.6}_{-2.6}$ & 35.1 \\
$h_0$ & $\mathcal{U}(0.005, 0.065)$ & $>0.028$ & 0.059 & $\mathcal{U}(0.005, 0.065)$ & $0.035^{+0.020}_{-0.020}$ & 0.06 \\
$\rm \alpha_{in}$ & $\mathcal{U}(1, 12)$ & $2.6^{+1.6}_{-0.8}$ & 2.7 & $\mathcal{U}(1, 12)$ & $>2.7$ & 11.7 \\
$\rm \alpha_{out}$ & $\mathcal{U}(-12, -1)$ & $-3.7^{+0.5}_{-0.7}$ & -3.5 & $\mathcal{U}(-12, -1)$ & $-1.3^{+0.2}_{-0.3}$ & -1.1 \\
$\rm \theta_{PA}$ (deg) & $\mathcal{U}(0, 180)$ & $164.7^{+5.6}_{-5.5}$ & 163.3 & $\mathcal{U}(0, 180)$ & $163.4^{+10.1}_{-10.8}$ & 158.9 \\
$\rm \theta_{inc}$ (deg) & $\mathcal{U}(0, 180)$ & $19.2^{+3.4}_{-2.7}$ & 17.0 & $\mathcal{U}(0, 180)$ & $19.6^{+8.3}_{-5.8}$ & 19.5 \\
$g_1$ & $\mathcal{U}(0, 1)$ & $>0.10$ & 0.94 & $\mathcal{U}(0, 1)$ & $>0.06$ & 0.98 \\
$g_2$ & $\mathcal{U}(-1, 0)$ & $-0.60^{+0.1}_{-0.1}$ & -0.68 & $\mathcal{U}(-1, 0)$ & $-0.55^{+0.21}_{-0.22}$ & -0.84 \\
$w_1$ & $\mathcal{U}(0, 1)$ & $<0.45$ & 0.1 & $\mathcal{U}(0, 1)$ & $<0.61$ & 0.24 \\
$\gamma$ & $\mathcal{U}(1, 3)$ & $2.1^{+0.6}_{-0.7}$ & 2.9 & $\mathcal{U}(1, 3)$ & $2.1^{+0.7}_{-0.7}$ & 2.5 \\
$\Delta x\ (\rm arcsec)$ & - & - & - & $\mathcal{U}(1, 2)$ & $1.304^{+0.025}_{-0.032}$ & 1.270 \\
$\Delta y\ (\rm arcsec)$ & - & - & - & $\mathcal{U}(-3.5, -2.5)$ & $-3.036^{+0.037}_{-0.036}$ & -2.979 \\
$f_{\rm bg}\ (\rm \mu Jy)$ & - & - & - & $\mathcal{U}(0, 5)$ & $1.36^{+0.24}_{-0.29}$ & 0.95 \\
\enddata
\tablecomments{Credible intervals are $68\%$ credible unless stated otherwise. Entries marked $>$ or $<$ are one-sided $95\%$ credible lower
or upper bounds, corresponding to the 5th or 95th posterior
percentile, respectively.}
\end{deluxetable*}

\section{Disk Forward Modeling} \label{sec:disks}

\subsection{MCMC Sampling}\label{sec:disk_model}

We used \texttt{Winnie} to perform forward modeling of the disk observed in both the F210M and F444W filters. The disk was modeled with the GRaTer framework \citep{Augerea1999}, assuming a single-ring structure in which the dust density $n(r,z)$ is described by radial and vertical components:

\begin{equation}
n(r,z) \propto R(r) Z(r,z),
\end{equation}
where $r$ is the radial distance from the star and $z$ is the height above the mid-plane. The radial distribution follows an asymmetric, broken power-law surface density profile defined by three parameters: the fiducial radius $r_0$, the density power law exponent interior to the fiducial radius $\alpha_{\mathrm{in}}$, and the density power law exponent exterior to the fiducial radius  $\alpha_{\mathrm{out}}$:

\begin{equation} R(r) \propto \left[ \left( \frac{r}{r_0} \right)^{-2\alpha_{\mathrm{in}}} + \left( \frac{r}{r_0} \right)^{-2\alpha_{\mathrm{out}}} \right]^{-1/2} \end{equation}
with $\alpha_{\mathrm{in}} > 0$ and $\alpha_{\mathrm{out}} < 0$. The vertical distribution is modeled as an exponential profile:

\begin{equation}
Z(z,r) \propto \exp\left[-\tfrac{1}{2} \left( \frac{z}{h_0 r} \right)^{\gamma} \right]
\end{equation}
where $h_0$ is the scale-height ratio at the fiducial radius. We adopt a uniform prior $h_0\sim \mathcal{U}(0.005, 0.065)$, corresponding to an allowed disk aspect ratio between $0.5\%$ and $6.5\%$ at the fiducial radius. $\gamma$ is the vertical density exponent: $\gamma = 2$ corresponds to a Gaussian profile, $\gamma = 1$ to a more exponential-like profile with broader wings, and $\gamma = 3$ to a profile more sharply peaked toward the midplane. We adopt a uniform prior, $\gamma \sim \mathcal{U}(1,3)$, to allow a range of plausible vertical profile shapes. The priors adopted for all model parameters are listed in Table~\ref{tab:rvs}.

We adopted a scattering phase function $p(\theta)$ described by a linear combination of two Henyey-Greenstein (HG, \citealt{Henyey1941}) functions as:

\begin{equation}
    p (\theta) = w_1 HG(g_1,\theta) + (1-w_1) HG(g_2,\theta)
\end{equation}
where $\theta$ is the scattering angle and $HG(g,\theta)$ is the single HG function. The scattering phase function is described by three free parameters: two HG asymmetry parameters $g_1$ and $g_2$, and the relative weight of the first HG component  $w_1$ ($0 \le w_1\le 1$). The first HG parameter $g_1$ ($0 \le g_1\le 1$) characterizes forward scattering, with values near 1 indicating strong forward scattering and values near 0 meaning isotropic scattering, while $g_2$ ($-1 \le g_1\le 0$) characterizes backward scattering, with values near $-1$ indicating strong backward scattering. 

In addition, the disk orientation is described by the position angle $\theta_{\rm PA}$ (measured east of north) and the inclination $\theta_{\rm inc}$. 
For each sampled disk morphology, we first convolved the synthetic image with the instrumental PSF and propagated it through the same RDI processing applied to the observations.  We then scaled the resulting forward model by a flux factor determined through least-squares minimization of the residuals within the region of interest. Thus, the disk flux normalization was optimized at each MCMC step rather than sampled as an independent parameter. This yields a total of 10 free parameters for the disk model. We assume the eccentricity and argument of pericenter as zero. For the F444W observations, we introduce three additional parameters to model the PSF of the source at southeast direction from the host star: its position relative to the central star in right ascension ($\Delta x$), its position in declination ($\Delta y$), and its flux ($f_{\rm bg}$). For each observation, we first rotate the  disk model
into the detector frame, apply the coronagraph transmission map, and convolve it with the instrumental PSF. The resulting models are then derotated into a common sky
orientation and combined using
the same procedure applied to the observed data.

To reduce the influence of spatial correlations between neighboring pixels, we binned the data and forward-modeled images in nonoverlapping $2\times2$ pixel bins after PSF subtraction and before evaluating the disk flux normalization and likelihood. Similar binning approaches have been used in scattered-light disk modeling \citep{Thalmann2014,MillarBlanchaer2016}. The likelihood is computed by comparing the observed disk image to the simulated model as:

\begin{equation}
\ln \mathcal{L} = -\frac{1}{2} \sum_{k=1}^{N} \frac{(\mathcal{M}_{k,\rm bin} - \mathcal{D}_{k,\rm bin})^2}{\sigma_{k,\rm bin}^2}
\end{equation}

where $\mathcal{M}_{k,\rm bin}$ and $\mathcal{D}_{k,\rm bin}$ are the mean model and observed fluxes in the $k$th bin, respectively, $\sigma_{k,\rm bin}$ is the uncertainty assigned to that bin, and $N$ is the number of bins included in the fit. For each bin, we assume that the noise among its four constituent pixels is fully correlated. The uncertainty of the mean flux in each bin is
\begin{equation}
\sigma_{k,\rm bin}
= \frac{1}{4}\sqrt{
\sum_{i=1}^{4}\sigma_i^2
+ 2\sum_{i<j}\rho_{ij}\sigma_i\sigma_j
},
\end{equation}
where $\sigma_i$ is the uncertainty of the $i$th constituent pixel and $\rho_{ij}$ is the noise correlation coefficient between pixels $i$ and $j$. Setting $\rho_{ij}=1$ gives
\begin{equation}
\sigma_{k,\rm bin} = \frac{1}{4}\sum_{i=1}^{4}\sigma_i.
\end{equation}

Note that binning reduces the influence of spatial correlations between neighboring pixels, but does not necessarily eliminate them. The likelihood assumes independent residuals between bins and does not explicitly account for any remaining spatial correlations; the resulting credible intervals may therefore underestimate the parameter uncertainties. To reduce computational time, we restrict the optimization region to within $4^{\prime\prime}$ of the star. We used the Markov chain Monte Carlo (MCMC) ensemble sampler in \textit{emcee} \citep{FM2013} to sample the parameter space with 50 walkers. We ran 20,000 steps per walker for F444W and 29,000 steps per walker for F210M, achieving \(N/\tau>50\) for every model parameter, where \(N\) is the chain length and \(\tau\) is the integrated autocorrelation time (see Appendix~\ref{fig:tau210} and ~\ref{fig:tau444}). This criterion indicates that the posterior distributions were adequately sampled. We discarded the initial $30\%$ of the chain as burn-in. Figure~\ref{fig:figure1} presents the observations, best-fit models and residuals. In F210M filter, we obtained a fiducial radius of $r_0=55.4^{+3.0}_{-3.2}$ AU, inclination of $\theta_{inc}=19^{\circ}.2 ^{+3.4}_{-2.7}$ and disk position angle $\theta_{\rm PA}=164.7^{+5.6}_{-5.5}$. In F444W filter, the derived fiducial radius is $r_0=39.1^{+4.6}_ {-2.6}$ AU, disk inclination is $\theta_{inc}=19^{\circ}.6 ^{+8.3}_{-5.8}$ and  position angle is $\theta_{\rm PA}=163.4^{+10.1}_{-10.8}$. Table~\ref{tab:rvs} presents the posterior medians and $1\sigma$ credible intervals, or $95\%$ credible limits where appropriate, for all free parameters in the model (see Appendix~\ref{fig:figureA210} and ~\ref{fig:figureA444} for joint posterior distributions). Although all free parameters pass our convergence checks, we note that the scale-height ratio at the fiducial radius $h_0$ and vertical density exponent $\gamma$ are only weakly constrained by the data. This is because the disk is observed close to face-on, so the vertical structure is poorly resolved in projection. However, the joint posterior distributions show no strong covariance between either $h_0$ or $\gamma$ and the disk inclination, indicating that the limited constraints on the vertical disk structure do not significantly affect the inferred inclination.


We find a discrepancy in the fitted fiducial radii ($r_0$) between the F210M and F444W observations, although their joint posterior distributions show marginal overlap at the $2\sigma$ level (see Figure~\ref{fig:corner_compare}). In contrast, the fitted disk inclinations and position angles (PAs) agree well between the two filters, with values consistent within their $1\sigma$ uncertainties . The larger fitted $r_0$ in F210M may arise because the shorter-wavelength observations are more sensitive to faint disk emission at larger separations, which shifts the apparent surface-brightness distribution outward. The discrepancy may also reflect limitations of the adopted simplified model, which represents the disk as a single ring despite the presence of extended emission. As discussed in Section~\ref{sec:resi}, the asymmetric residuals in the F210M image suggest that this model does not fully capture the observed morphology, and may therefore bias the inferred 
$r_0$ toward larger values. Furthermore, the dust-scattering phase function would be  wavelength-dependent, therefore cause to different observed surface-brightness distribution in F444W and F210M filters. The difference may also be influenced by PSF-subtraction systematics, which can distort the recovered surface-brightness profile differently in the two filters. However, we find that the discrepancy in $r_0$
does not propagate into the inferred disk inclination. The joint posterior distributions show no strong covariance or systematic correlation between $r_0$ and inclination. More directly, although the two filters yield different characteristic radii, they independently recover consistent disk inclinations. This agreement indicates that, despite differences in the inferred radial structure between the two datasets, the recovered projected disk geometry remains consistent. 


\begin{figure*}
    \centering
    \includegraphics[width=\linewidth]{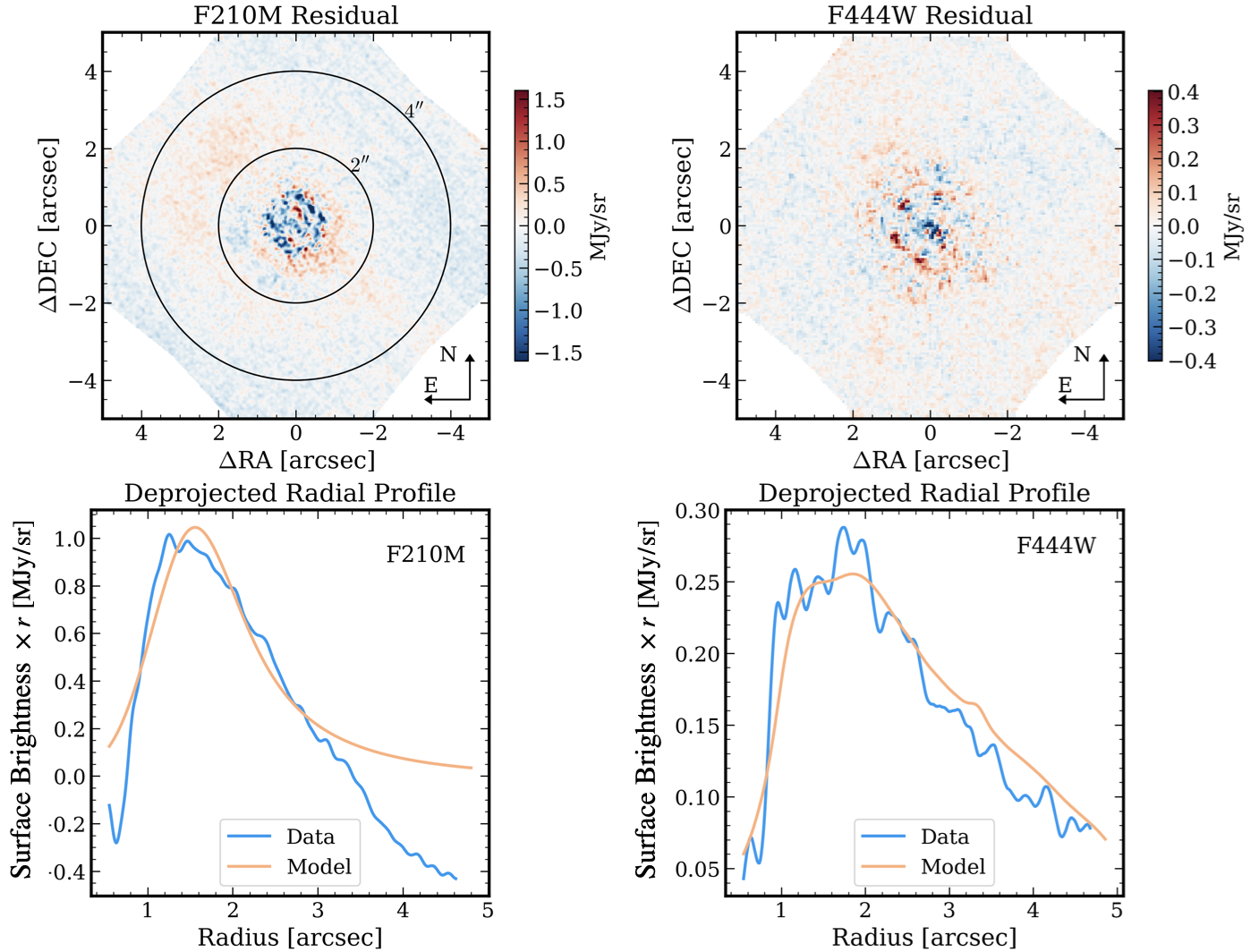}
    \caption{(a) $\&$ (b): residuals after subtracting best-fit disk model in F210M and F444W. (c) $\&$ (d): the de-projected radial profile of observed disk (blue) and best-fit model (peach) in F210M and F444W. The y-axis represents the intensity of an annulus inside radius of $r$ with an interval of $\Delta r = 0.1^{\arcsec}$, multiplied by the radius.  }
    \label{fig:figure_residual}
\end{figure*}

\subsection{Comparison with ALMA and PACS observations}\label{sec:resi}


The disk around CPD-72 2713 was previously resolved in thermal emission with ALMA at 1.33 mm and the Photodetector Array Camera and Spectrometer (PACS) on the Herschel Space Observatory at  $100 \mu m$ and $160 \mu m$.  \cite{Tanner2020} analyzed the Herschel/PACS observations and reported a disk radius of $96\pm{7.4}$ AU and disk inclination of $7^{\circ}\pm{19}$, which agrees with our JWST measurements within $1\sigma$.  \citet{Moor2020} modeled the disk using the low-resolution ALMA dataset (7 m array, beam size $7^{\arcsec}.4 \times 5^{\arcsec}.5$), obtaining disk radii of $R = 140 \pm 14$ AU, disk inclination of $i=46^{\circ}\pm{12}$ and position angle (PA) of $\rm PA= 130^{\circ}\pm{6}$. More recently, \citet{Marino2026} presented high-resolution ALMA observations (beam size $0.55^{\arcsec}$) of the system and reported a disk radius of $95.7\pm2.5$ AU, an inclination of $18^{\circ}\pm8^{\circ}$, and a PA of $149^{\circ}\pm20^{\circ}$. \citet{Han2026} analyzed the same high-resolution ALMA dataset and inferred a disk radius of $98^{+6}_{-7}$ AU, an inclination of $32^{+3}_{-4}$ degrees, and a PA of $147^{+8}_{-16}$ degrees. The two analyses adopted different modeling approaches: \citet{Marino2026} assumed a symmetric Gaussian surface-density profile and simultaneously fitted the disk geometry, the alignment of the individual ALMA integrations , and the background source, whereas \citet{Han2026} adopted a more flexible asymmetric surface-density profile and fitted the aligned and background-subtracted data with the background-source parameters fixed. A detailed assessment of the different modeling approaches and their effects on the inferred ALMA disk geometry is beyond the scope of this work. We therefore present both ALMA measurements in our comparison. Our independently derived JWST inclination agree well with value reported by \citet{Marino2026}, but differs from that of \citet{Han2026} beyond the $1\sigma$ level, although their $2\sigma$ credible intervals overlap. These comparisons should be interpreted in the context of the modeling assumptions. Our NIRCam credible intervals are conditional on the adopted disk model and likelihood. Although \(2\times2\) pixel binning reduces the influence of correlations between neighboring pixels, the likelihood assumes independent residuals between bins. Remaining spatial correlations and model-dependent uncertainties may therefore still lead to underestimated uncertainties \citep{Wolff2017,Czekala2015}. The ALMA inclinations were derived from visibility fitting and are not subject to the same reconstructed-image pixel covariance, although they remain conditional on the adopted noise and disk models. 

In terms of disk radius, compared to the JWST/NIRCam scattered-light images in the near-infrared ( disk radius $r_0\sim55.4\ \mathrm{AU}$ in F210M and $r_0\sim 39.1 \mathrm{AU}$ in F444W), the ALMA and PACS observations reveal a much more extended radial structure, reaching $r \sim 100\ \mathrm{AU}$. One interpretation is that  the three instruments are sensitive to different grain populations: ALMA/PACS measures thermal emission from cold dust distributed across large spatial scales, while JWST detects scattered starlight, which primarily traces small, efficiently scattering grains. The scattered light brightness declines with distance as $1/r^2$, and therefore JWST/NIRcam is less sensitive to faint structures at larger distances. This is aligned with recent studies from JWST MIRI observations showing that small grains traced at shorter wavelengths can appear substantially interior to the larger-grain planetesimal belts traced at millimeter wavelengths \citep{Gaspar2023,su2024, wolf2025}.  \cite{sommer2025} also points out that this behavior may arise from size-dependent dust dynamics, including inward transport by Poynting–Robertson drag. 

\begin{figure*}
    \centering
    \includegraphics[width=\linewidth]{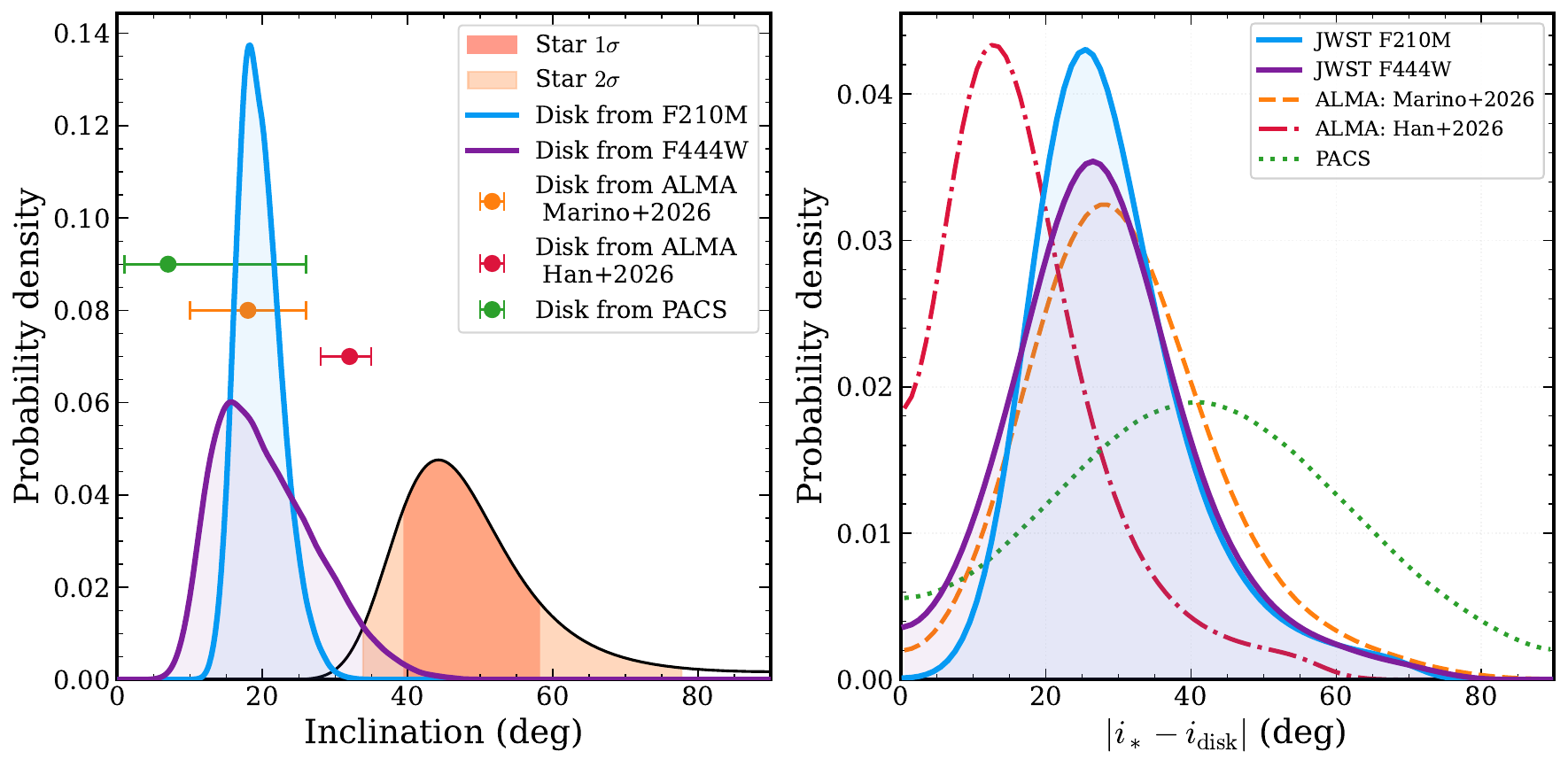}
    \caption{ Left: Posterior distribution of the stellar inclination compared with disk inclination measurements from different datasets. The dark and light orange bands denote the $1\sigma$ and $2\sigma$ credible regions of stellar inclination. The blue and purple histograms show the posterior distributions of disk inclination from the JWST/NIRCam F210M and F444W filters. The orange, red and green points indicate disk inclinations reported in \cite{Marino2026}, \cite{Han2026} and \cite{Tanner2020}, respectively. Right: the distribution of the minimum star-disk mutual inclination from JWST/NIRCam F210M (blue), JWST/NIRCam F444W (purple), ALMA (orange,red) and PACS (green) observations. Each distribution was constructed by randomly drawing from the disk- and stellar-inclination distributions and smoothing the resulting histogram with a Gaussian filter.  }     
    \label{fig:figurestellar_inc}
\end{figure*}

\subsection{Residual in F210M filter}

Figures~\ref{fig:figure_residual}a and b show the residuals after subtracting the disk model from the JWST/NIRCam observations. The residuals in F210W show an intensity excess in the northeastern region, between $2^{\arcsec}$ and $4^{\arcsec}$, and an excess in the opposite direction, towards the southwestern region, but at a much closer distance to the star ($1^{\arcsec}$). To further investigate this feature, we tested the disk model in F210M by adding a second ring or halo component, but this did not result in a better fit than the single-ring model. The de-projected radial profiles also do not show a distinct second ring or bump associated with the halo (see Figures~\ref{fig:figure_residual}c and d). Meanwhile, we observed that the asymmetry aligns with the minor axis of the disk, suggesting that the residual excess may be from forward scattering from an non-axisymmetric disk. One possible interpretation is that the phase function varies with the disk radius. However, our disk model, which uses a double power-law radial profile combined with two HG phase functions, is not flexible enough to capture that. As a result, the model is attempting to find a compromise between fitting the inner and outer regions. On the other hand, we do not observe a significant excess or asymmetry in the residuals of the F444W observations.
An background galaxy could also contribute to the F210M excess, although the current data do not distinguish this possibility from PSF-subtraction residuals or disk structure not captured by our model.

Another plausible mechanism for the observed emission asymmetry is a non-zero disk eccentricity. To test this, we also performed disk-model fitting to the F210M data with eccentricity included as a free parameter. We obtained a value consistent with zero,$e=0.014^{+0.016}_{-0.011}$, indicating no evidence for a significantly eccentric disk. It is therefore unlikely that disk eccentricity is responsible for the observed asymmetry. We note the presence of negative emission outside the disk in the F210M data, which likely arises from the PSF-subtraction process. Our model does not explicitly account for these negative emission, and we acknowledge this as a limitation of our model.


\section{Stellar-Disk Misalignment}\label{sec:sd_misalign}

\subsection{Stellar Inclination}
We measured a stellar rotation period of $P_\mathrm{rot} = 4.47 \pm 0.67$ days using 2-minute cadence TESS SPOC light curves in sectors 1, 13, 27, 28, 67, 68, and 94 (Figure~\ref{fig:figurerot}). The uncertainty includes the width of the periodogram peak and an additional $10\%$ contribution for surface differential rotation \citep{Epstein2014, Claytor2022}. The TESS Input Catalog reports a low contamination ratio of 0.00318 for CPD−72 2713 \citep{tessdata}. Its Gaia RUWE of $\sim 0.93$ is consistent with a well-behaved single-star astrometric solution \citep{Gaia}, and our JWST imaging rules out stellar companions at angular separations of $\sim$0.1--10 arcsec. These results support attributing the detected photometric modulation and inferred rotation period to CPD−72 2713.

Furthermore, the projected stellar rotational velocity, $v\sin i_\ast$, has been measured using high-resolution stellar spectra. Observations using FEROS \citep{Kaufer1999} on the MPG/ESO 2.2 m telescope obtained  $7.3\pm1.0~\mathrm{km\ s^{-1}}$ \citep{Zakhozhay2022} and $7.5\pm1.2~\mathrm{km\ s^{-1}}$ \citep{Torres2006}, respectively. Measurements obtained with HARPS \citep{Mayor2003} on the ESO 3.6 m telescope at La Silla yielded $6.6\pm1.4~\mathrm{km\ s^{-1}}$ \citep{Weise2010}. Spectra taken with Ultraviolet and Visual Echelle Spectrograph \citep{Dekker2000} in Paranal, Chile obtained $5.9\pm0.4~\mathrm{km\ s^{-1}}$ \citep{ZF2021}
All four measurements agree within their quoted $1\sigma$ uncertainties. Rather than selecting a single measurement, we follow \cite{Bowler2023} and adopt the weighted mean of the four measurements and impose a minimum fractional uncertainty of $15\%$. We obtained a weighted mean value of $v\sin i_\ast = 6.7 \pm 0.6~\mathrm{km,s^{-1}}$. We adopt the stellar radius from \cite{Marino2026}. CPD-72 2713 is a young, magnetically active star, for which unmodeled stellar spots may bias the inferred stellar radius. To account for this potential systematic effect, we adopt an additional $10\%$ uncertainty.
By combining the measurements of $P_\mathrm{rot}$ and stellar radius $R_*$ , we can derive the equatorial rotational velocities of the star as $v_{eq}=2\pi R_* /P_\mathrm{rot}$. If $R_*$, $P_{\rm{rot}}$ and $v\sin{i_*}$ are independent, $i_*$ can be estimated as $\sin^{-1}(\frac{ v\sin{i_*}}{v_{eq}})$. However, \cite{MandW2020M} showed that $v_{\mathrm{eq}}$ and $v\sin i_\ast$ are not statistically independent, so their correlation must be taken into account when deriving the probability distribution of $i_\ast$.  To account for the correlation between $v\sin i_\ast$ and $v_{\mathrm{eq}}$, \cite{Bowler2023} presented a Bayesian probabilistic framework to constrain the stellar inclination (see Appendix A of \citealt{Bowler2023}) as 
 
\begin{equation}{\label{eqn:i}}
P(i_*\mid P_\mathrm{rot}, R_*, v\sin i_*) \propto  \sin i_* \times \frac{e^{- \frac{\big(v \sin i_* - \frac{2\pi R_*}{P_\mathrm{rot}}\sin i_* \big)^2}{2\big(\sigma_{v\sin i_*}^2 + \sigma_{v_\mathrm{eq}}^2 \sin^2 i_* \big)}} }{\sqrt{\sigma_{v\sin i_*}^2 + \sigma_{v_\mathrm{eq}}^2 \sin^2 i_*}}
\end{equation}
where $\sigma_{v\sin i_*}$ and $\sigma_{v_\mathrm{eq}}$ are the uncertainties in the projected velocity $v\sin{i_{*}}$ and the equatorial velocity $v_{eq}$. We approximated $\sigma_{v_\mathrm{eq}}$ as
\begin{equation}{\label{eqn:veqerr}}
\sigma_{v_\mathrm{eq}} =  \frac{2\pi R_*}{P_\mathrm{rot}} \sqrt{\Big(\frac{\sigma_{R_*}}{R_*}\Big)^2 + \Big(\frac{\sigma_{P_\mathrm{rot}}}{P_\mathrm{rot}}\Big)^2 }
\end{equation}
 
Figure~\ref{fig:figurestellar_inc}a shows the resulting distribution for stellar inclination of CPD-27 7213. Using the adopted mean-weighted $v\sin{i_*}$, we obtained a stellar inclination  as $i_{\star}=47.2^{+11.1}_{-7.7}\ (1\sigma)$.
Our result is consistent with the values $42^{\circ}\pm{7}$ reported by \cite{Moor2020} \added{using the same method}. To assess the sensitivity of the derived stellar inclination to the adopted $v\sin i_*$, we also calculated the stellar inclination using four independent $v\sin i_*$ measurements from \cite{Torres2006,Weise2010,ZF2021,Zakhozhay2022}, in addition to our adopted value. The resulting stellar inclinations are  $50^{\circ}.2^{+16.9}_{-10.8}$, $45^{\circ}.4^{+18.6}_{-12.5}$, $41^{\circ}.8^{+10.8}_{-8.0}$, and $52^{\circ}.6^{+19.4}_{-11.7}$, respectively. These values are all consistent with one another within their $1\sigma$ uncertainties. The four inferred stellar inclinations have a mean of $47^{\circ}.5$ and a sample standard deviation of $\sim 4^{\circ}$, substantially smaller than their individual $1\sigma$ uncertainties ($8^{\circ}\sim19^{\circ}$), indicating that the variation introduced by adopting different published $v\sin{i_*}$ measurements is smaller than the quoted uncertainty of the inferred stellar inclination.


\subsection{Star-disk mutual inclination}

In this work, we define the stellar--disk mutual inclination as the angle between the stellar spin axis and the normal vector to the disk plane. A mutual inclination of $0^{\circ}$ corresponds to perfect alignment, while $90^{\circ}$ corresponds to orthogonality. Because we can constrain only the inclination of the stellar spin axis relative to the line of sight, while its orientation in the sky plane remains unknown, the true three-dimensional stellar--disk mutual inclination cannot be determined. Instead, the absolute difference between the stellar and disk inclinations provides a lower limit on the mutual inclination, which we refer to as the minimum mutual inclination.

Figure~\ref{fig:figurestellar_inc} compares the stellar inclination with the disk inclinations derived from different datasets. The left panel shows the stellar inclination distribution, described by Eqn.~\ref{eqn:i}, together with the disk inclination constraints. Our JWST scattered-light observations show an offset between the disk and stellar inclinations, indicating a misalignment between the disk traced by JWST and the stellar equatorial plane.

The right panel of Figure~\ref{fig:figurestellar_inc} shows the corresponding minimum star-disk mutual inclination distributions. We construct these distributions by drawing 50,000 random samples from the stellar and disk inclination distributions and computing the absolute difference between the two inclinations for each draw. For the JWST observations, the disk inclination samples are drawn directly from the posterior distributions obtained from our modeling of the two filters. For the ALMA and PACS measurements, we draw samples from distributions based on their reported inclination values and uncertainties. 

For the JWST observations, we obtain minimum star--disk mutual inclinations of $27^{\circ}.6 _{-8.2} ^{+11.5}$ ($1\sigma$) in F210M and $27^{\circ}.3 _{-10.9}^{ +12.5}$ ($1\sigma$) in F444W. The lower bounds of the  $95\%$  credible intervals are $14.7^\circ$ and $10.2^\circ$ for F210M and F444W, respectively. Therefore, both filters indicate a minimum mutual inclination exceeding $10^\circ$ at the $95\%$ credible level, providing evidence that the disk traced in scattered light is misaligned with the stellar equatorial plane. Because the difference between the stellar and disk inclinations represents only a lower limit on their full three-dimensional mutual inclination, the true star--disk misalignment may be larger.

For reference, we also discuss the star-disk mutual inclination in the context of the disk inclinations reported in previous studies. \citet{Moor2020} previously reported an alignment between the disk and the star based on lower-resolution ALMA observations. However, their inferred disk inclination may be affected by the limited spatial resolution of those data. In particular, the low-resolution observations do not resolve the bright background source identified in the later high-resolution ALMA data, and its flux may therefore have been attributed to the disk in the modeling.
We therefore focus our comparison on the more recent disk inclination measurements derived from the high-resolution ALMA data. The disk inclination reported by \citet{Marino2026} is in good agreement with our JWST measurement and implies a similar degree of star-disk misalignment. In contrast, the higher disk inclination reported by \citet{Han2026} would imply a smaller minimum star--disk mutual inclination and therefore less significant evidence for misalignment. As discussed in Section~\ref{sec:resi}, \citet{Marino2026} and \citet{Han2026} adopted different modeling approaches, and assessing the origin of the difference between their inferred disk inclinations is beyond the scope of this work.

 \begin{figure*}
    \centering
    \includegraphics[width=\linewidth]{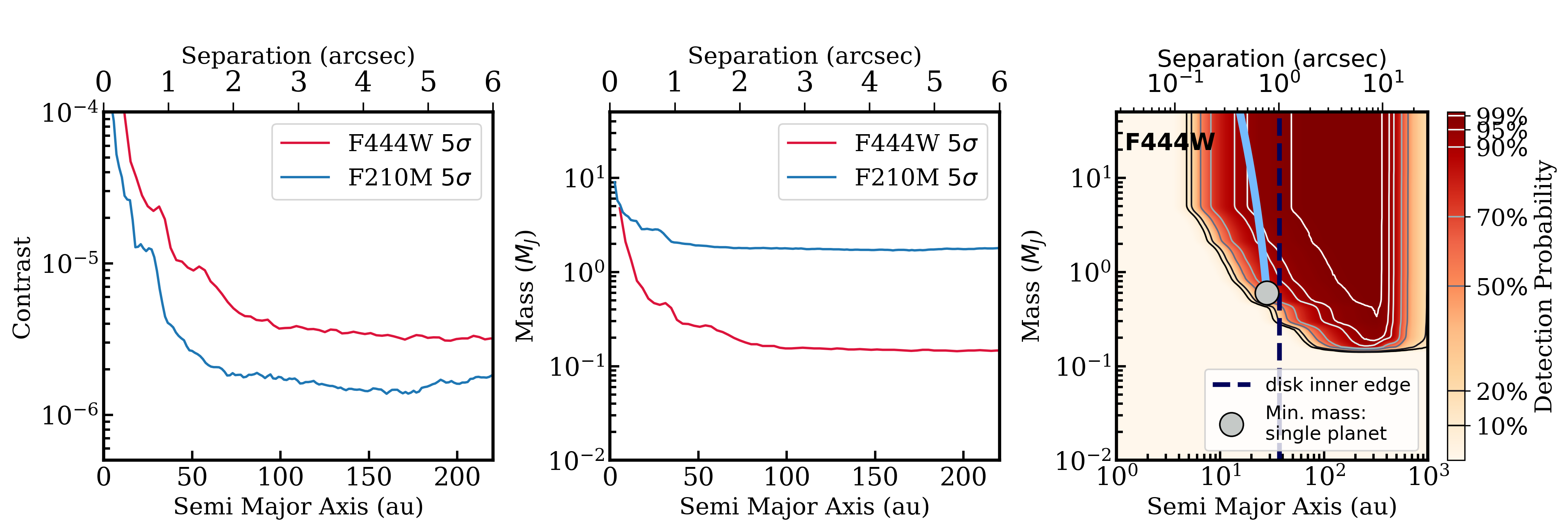}
    \caption{Left: Contrast curves ($5\sigma$) as a function of projected separation after MCDRI processing and disk subtraction for the F210M (blue) and F444W (red) filters. Middle: Mass sensitivity limits as a function of projected separation for the F210M (blue) and F444W (red) filters. The F444W filter is sensitive to lower planetary masses than the F210M filter despite achieving worse contrasts due to the fact that planets are often much brighter at the wavelengths of the F444W filter. Right: Detection probability map generated by \textit{maday} for the F444W filter. The vertical dashed line marks the disk inner edge observed in F444W, while the blue line shows the predicted mass and separation of a single sculpting planet responsible for the observed inner edge, based on the dynamical model of \cite{Pearce2022}. The gray circle indicates the minimum planet mass required by the model. }
    \label{fig:figure_contrast}
\end{figure*}



\section{JWST/NIRCam direct imaging detection limit}
\subsection{Search for Companions} \label{sec:plantes}

We searched for planet companions in the system of CPD -27 7213 in the disk-subtracted images. As discussed in Section~\ref{sec:disks}, we identified a source at approximately $\Delta\mathrm{RA} \sim 1.3^{\prime\prime}$ and $\Delta\mathrm{Dec} \sim -3.2^{\prime\prime}$ in the F444W filter, but no counterpart is detected in the F210M filter. To characterize it, we performed simultaneous forward-modeling of the disk and companion psf to the reduced RDI data using MCMC with \texttt{Winnie} in section~\ref{sec:disk_model}. Results of the companion modeling are summarized in Table~\ref{tab:rvs}. 

\cite{Marino2026} also detected a source in the similar location ($\Delta \rm{RA} = 1.45^{\arcsec}\pm{0.70}$, $\Delta \rm{DEC}= -3.05^{\arcsec}\pm{0.60}$ ) from their high-resolution ALMA observations taken at Oct. 2022. We used Package \texttt{backtracks} \citep{william_o_balmer_2025_14838370} to perform a common-proper-motion test with two astrometric epochs spanning $\sim$2 years (see Figure~\ref{fig:figureback}). We found the source follows the expected background track. Combined with the non-detection at blue wavelengths (F210M), we conclude the source is most likely a background galaxy rather than a background star.



\subsection{Contrast curves and detection probability}
Figure~\ref{fig:figure_contrast} left panel presents the planet-star flux ratio as a function of angular separation, i.e. contrast curves, for the observations in both F210M and F444W filters. To avoid biasing the contrast measurements due to disk emission, we used \texttt{Winnie}, which implements the contrast calculation functionality from \texttt{spaceKLIP}, on images with the best-fit disk model subtracted. In F444W observations, we also masked out the background galaxy.
Since algorithmic throughput losses are negligible for MCRDI, we follow \cite{Lawson2024,Crotts2025} and adopt raw contrast curves from \texttt{SpaceKLIP} as the final contrasts for our MCRDI reductions. These raw contrasts already account for the coronagraph transmission. The stellar flux was estimated in \texttt{SpaceKLIP} using the input stellar photometry from \textit{Spitzer} \citep{spitzerdata}, \textit{ALLWISE} \citep{allwisedata}, ASAS-SN \citep{asasdata}, and \textit{TESS} \citep{tessdata} surveys. 

Figure~\ref{fig:figure_contrast} middle panel shows the sensitivity of JWST NIRCam to different planetary masses in the CPD -72 2713 system based on the $5\sigma$ contrast curves. We used the Python package \texttt{species} \citep{stolker2020miracles} and adopted the evolutionary models from \citet{Linder2019} and \citet{Chabrier2023}, selecting the cloud-free case with nominal post-formation luminosity and solar metallicity ($[M/H] = 0$). Although the achieved contrast is shallower in F444W, the mass sensitivity is better because planets are intrinsically brighter at $4 \mu\mathrm{m}$ than at $2 \mu\mathrm{m}$. In the F210W filter, NIRCam is sensitive to masses down to 2 $\rm M_J$ beyond  $0.5^{\arcsec}$  separation; for the F444W filter, NIRCam is sensitive to masses down to 0.3 $\rm M_J$ beyond $0.5^{\arcsec}$ separation. 

Figure~\ref{fig:figure_contrast} right panel presents the detection probability map for the F444W filter. We used the Python package \texttt{MADYS} \citep{Squicciarini2022} to compute the map from the contrast curve. Specifically, we injected a population of synthetic companions with masses between $0.01,M_{\rm J}$ and $50,M_{\rm J}$ and semimajor axes between 1 and 1000 AU. Their eccentricities were drawn from a half-normal distribution with a width of 0.3 \citep{Hogg2010}, inclinations were assumed to be uniformly distributed in $\cos i$, and all other orbital elements were drawn from uniform distributions. To convert companion brightness into mass, we adopted the evolutionary models of \citet{Linder2019}. 

\citet{Pearce2022} argued that, if one single planet interior to the disk truncate the disk edge, then there should either be one planet with mass greater than ${1.1^{+0.3}_{-0.2} \; \rm M_{J}}$ inwards of ${60\pm20 \; \rm au}$. However, their predictions were based on an inner edge at 80 AU, inferred from partially resolved ALMA observations \citep{Moor2020}. In contrast, our JWST/NIRCam data clearly reveal an inner disk component within 80 AU. We recalculate the predicted planet properties  using our system values from \mbox{Table \ref{tab:star-compare}} and our new F444W disk fit from \mbox{Table \ref{tab:rvs}}, using the {\sc SculptingPlanet} online tool\footnote{ We used the publicly available SculptingPlanet code at Git commit 301f1af. The GitHub repository is available at \url{https://github.com/TimDPearce/SculptingPlanet}} \citep{Pearce2022}. In the model, the planet clears material within an instability region extending to 5 Hill radii, so matching the edge of this region to the observed disk edge can provide the constraint on the the sculpting planet's properties.  Because the Hill radius depends on both planet mass and semi-major axis, the inferred planet is degenerate in these two parameters: a more massive planet located farther from the disk can truncate the disk edge at the same location as a lower-mass planet orbiting closer to it. Requiring that the planet clear the debris within the system age further sets a minimum planet mass, corresponding to the solution located closest to the disk.  The resulting  constraint on the mass and semi-major axis is shown as the blue curve overplotted on the detection probability map, and the gray circle marks the minimum planet mass required by the model. We obtained that, if the disk inner edge is sculpted by a single planet, then that planet has a mass of at least ${0.62^{+0.06}_{-0.05} \; \rm M_{J}}$ and is located within ${29^{+3}_{-2} \; \rm au}$ (${0.79^{+0.08}_{-0.05} \; \rm arcsec}$). The detection probability at this minimum mass is approximately $70\%$, and and increases for more massive planets along the blue curve. Therefore, if a single planet is sculpting the inner edge, our observations would have a greater than $70\%$ probability of detecting it. \cite{Pearce2022} also considered a scenario in which multiple equal-mass planets on circular orbits occupy the region interior to the debris disk and clear that region within the system age. Under this scenario, each planet would have a minimum mass of $0.05 \pm 0.01~\mathrm{M}_{\mathrm{J}}$ using the disk parameters of CPD -72 2713, which are are below the detection limit of the JWST observations from \mbox{Figure \ref{fig:figure_contrast}}.

\section{Discussion}\label{sec:diss}

\subsection{How can the disk become misaligned relative to the host star?
}


In the Solar System-inspired picture of planet formation, planets form in a flat protoplanetary disk aligned with the stellar equator, yielding orbits that are nearly circular and coplanar. However, exoplanet observations show that alignment is not universal. The  misalignment between stellar spin axis and orbits of hot Jupiters have long been detected via the Rossiter-McLaughlin effect \citep{R1924, McLaughlin1924} or Doppler shadows \citep{Albrecht2007}, particularly around hot stars with effective temperatures above the Kraft break ($\sim\!6250$\,K; \citealt{Winn2010,Schlaufman2010,Albrecht2012}). Spin-orbit misalignments are also observed in systems hosting small planets ($R_p<4\,R_{\oplus}$) \citep{Huber2013,Hjorth2021,zhang2021}, and mutual inclinations among planets orbiting the same star are increasingly reported \citep{Xuan2020,Zhang2025a}. There two possible  mechanisms responsible for these observed misalignments: (1) primordial misalignment in the natal disks where planets form and (2) post-formation misalignment caused by the gravitational interaction between planets, such as planet-planet scattering \citep{chatterjee2008}.

In the primordial scenario, disk could be misaligned relative to the host stars in the early stage of the planetary history. Using stellar rotation and $v\sin{i_*}$, \cite{Hurt2023} constrained the stellar inclination of 31 systems hosting resolved debris disks and found 6 among them with star-disk mutual inclination at least $30^{\circ}$ to $60^{\circ}$. With the same method,  \citet{Biddle2025} report that roughly one-third of Sun-like stars host protoplanetary disks that are misaligned relative to the stellar spin axis. In addition, high-resolution ALMA observations of disks have also revealed disk may not be uniform and flat, but with warps and misalignment between inner and outer disks \citep{Nealon2019}. Our observations show that CPD -27 2713 joins the small but growing list where the debris disk become misaligned relative to the host stars.

Several mechanism has been proposed to misalign the disk relative to the host star. One possibility is through secular processes disturbed by an inclined stellar or planetary companion \citep{Lubow2000, Batygin2012}.  If the star, planet, and debris disk were initially aligned with each other, a later dynamical event such as planet-planet scattering could have tilted the planet's orbit relative to the stellar spin axis and the debris disk. The inclined planet would then drive nodal precession of the disk planetesimals, producing a warp that propagates outward and gradually reorients the disk toward the planetary orbital plane, thereby generating a misalignment between the debris disk and the star \citep{Mouillet1997, Sefilian2025}.  If we assume the disk is massless, the time for the warp to reach the outer edge of the belt, and thus for the whole disk to become approximately aligned with the planet and therefore misaligned with the star, is given by \citep{ Sefilian2025}
\begin{equation}
t_{\rm align} \approx \frac{1}{2}T_{\rm sec}(a_{\rm out})
= \frac{\pi}{|A_p(a_{\rm out})|}
\end{equation}
where
\begin{equation}
A_p(a) = -\frac{1}{4}n(a)\frac{m_p}{M_\star}\frac{a_p}{a}
\,b_{3/2}^{(1)}\!\left(\frac{a_p}{a}\right)
\end{equation}
Using the detection limit from our JWST observation (\(m_p=0.3\,M_J\), \(a_p=18\) au($\sim0.5^{\arcsec}$)), and \(a_{\rm out}=100\) AU, we obtain \(t_{\rm align}\approx 57\) Myr, longer than the $\sim 24$ Myr age of the system. A lower-mass planet and/or a planet at smaller orbital separation would act even more slowly, with \(t_{\rm align}\propto m_p^{-1}a_p^{-2}\) in the \(a_p \ll a_{\rm out}\) limit. We therefore conclude that secular perturbation from an inclined planetary companion is unlikely to explain the observed star-disk misalignment in this system.

Another mechanism is the chaotic accretion, during which the late arrival of material from the molecular cloud warps or tilts the disk \citep{Bate2010,Thies2011}. Furthermore, the interaction between a magnetic star and its circumstellar disc could also push the stellar spin axis away from the disc angular momentum axis \citep{lai2011}. In addition, a stellar flyby may also tilt the orientation of the disk \citep{Moore2020}. Our observation can not distinguish  which, if any, of these mechanisms is responsible, but our data, combined with ALMA images, makes it clear that the CPD-72 2713 disk is tilted relative to its star.

An important question is whether stellar--disk misalignment is inherited by planetary orbits, producing planets that remain aligned with the disk but are misaligned with the stellar equator. Measurements of young planetary systems provide complementary constraints on their early orbital architectures. For example, the $\sim45$ Myr-old planet DS Tuc A b has a low sky-projected stellar obliquity \citep{Zhou2020,Montet2020}, indicating the planet is aligned with the host star. However, planetary orbits do not necessarily share the orientation of the outer disk. \citet{Barber2024} discovered a transiting planet around the $\sim3$ Myr-old star IRAS 04125+2902 whose nearly edge-on orbit is substantially misaligned with its outer protoplanetary disk. The stellar spin inclination is also consistent with an edge-on orientation, suggesting possible stellar-planetary alignment despite the tilted outer disk.



\section{Conclusion}\label{sec:conc}

In this work, we presented the JWST/NIRCam imaging
of the debris disk system, CPD-72 2713, in the F210M and F444W filters.  We detailed the results of the forward modeling of the observed disk and the measurement of the star-disk mutual inclination:

\begin{itemize}
    \item We detected the debris disk in scattered-light images from both filters, with the disk appearing brighter at shorter wavelengths (F210M) than at longer wavelengths (F444W). Using forward modeling with \texttt{Winnie} and MCMC sampling, we derived a disk radius of $\rm 55.4^{+3.0}_{-3.2}\,AU$ and an inclination of $19^{\circ}.2^{+3.4}_{-2.7}$ in F210M, and a radius of $\rm 39.1^{+4.6}_ {-2.6}\,AU$ and an inclination of $19^{\circ}.6^{+8.3}_{-5.8}$ in F444W. These scattered-light radii are significantly smaller than those inferred from thermal emission observed with ALMA ($95.7\pm{2.5}$ AU, \citep{Marino2026}) and Herschel/PACS ($96\pm{7.4}$, \citealt{Tanner2020}). One possible explanation is that different wavelengths probe different radial components of the disk: scattered light in the near-infrared is more sensitive to the inner regions, whereas thermal emission traces the cooler, outer disk. 

    \item We identified a stellar rotation period of $4.47 \pm 0.67$ days using seven \textit{TESS}-sector light curves. Combining this with the spectroscopically measured stellar rotation velocity ($v \sin i_*$), we constrained the stellar inclination to $i_* = 47^{\circ}.2^{+11.1}_{-7.7}$. Comparing this with the disk inclination derived from our modeling, we find that the disk seen in JWST scattered-light images is misaligned with the stellar equatorial plane by at least $13^{\circ}.6$ and $7^{\circ}.8$ in the F210M and F444W observations  with $99.7\%$ credible level. The star-disk mutual inclination in line-of-sight direction are $27^{\circ}.6 _{-8.2} ^{+11.5}\ (1\sigma)$ in F210M observations  and $27^{\circ}.3 _{-10.9}^{ +12.5}\ (1\sigma)$ in F444W observations, respectively. These constraints are conditional on the adopted disk models and likelihood assumptions. Our JWST disk inclinations are consistent with the measurements from \cite{Marino2026} and Herschel/PACS, while the higher inclination reported by \cite{Han2026} only overlaps our estimates at the $2\sigma$ level.

    \item We detected a source at $\Delta \mathrm{RA} = 1.304^{\prime\prime}$ and $\Delta \mathrm{Dec} = -3.036^{\prime\prime}$ in the F444W filter. Combined with high-resolution ALMA observation, we conclude that this object is a background galaxy.

\end{itemize}

\acknowledgments

This work is based on observations made with the NASA/ESA/CSA James Webb Space Telescope. The data were obtained from the Mikulski Archive for Space Telescopes at the Space Telescope Science Institute, which is operated by the Association of Universities for Research in Astronomy, Inc., under NASA contract NAS 5-03127 for JWST. These observations are associated with program 6012 (PI: M. A. Millar-Blanchaer). Support for program 6012 was provided by NASA through a grant (JWST-GO-06012.001-A) from the Space Telescope Science Institute, which is operated by the Association of Universities for Research in Astronomy, Inc., under NASA contract NAS 5-03127.

TDP was supported by a UKRI Stephen Hawking Fellowship and a Warwick Prize Fellowship, the latter made possible by a generous philanthropic donation.

E.C., L.A., and N.G. ackowledge funding from the European Union (ERC, ESCAPE, project No 101044152). Views and opinions expressed are however those of the author(s) only and do not necessarily reflect those of the European Union or the European Research Council Executive Agency. Neither the European Union nor the granting authority can be held responsible for them.

SM acknowledges funding by the Royal Society through a Royal Society University Research Fellowship (URF-R1-221669) and the European Union through the FEED ERC project (grant number 101162711). 

This work benefited from the 2025 Exoplanet Summer Program in the Other Worlds Laboratory (OWL) at the University of California, Santa Cruz, a program funded by the Heising-Simons Foundation.

\vspace{5mm}
\facilities{JWST(NIRCam), TESS}


\software{spacekilp \citep{Kammerer2022, Carter2023},  
          jwst \citep{Bushouse2022}, 
          winnie\citep{Lawso2022,Lawson2023}, madys\citep{Squicciarini2022}
          }



\newpage
\counterwithin{figure}{section}
\appendix
\section{Supplementary figures}
\begin{figure}[!b]
    \centering
    \includegraphics[width=\linewidth]{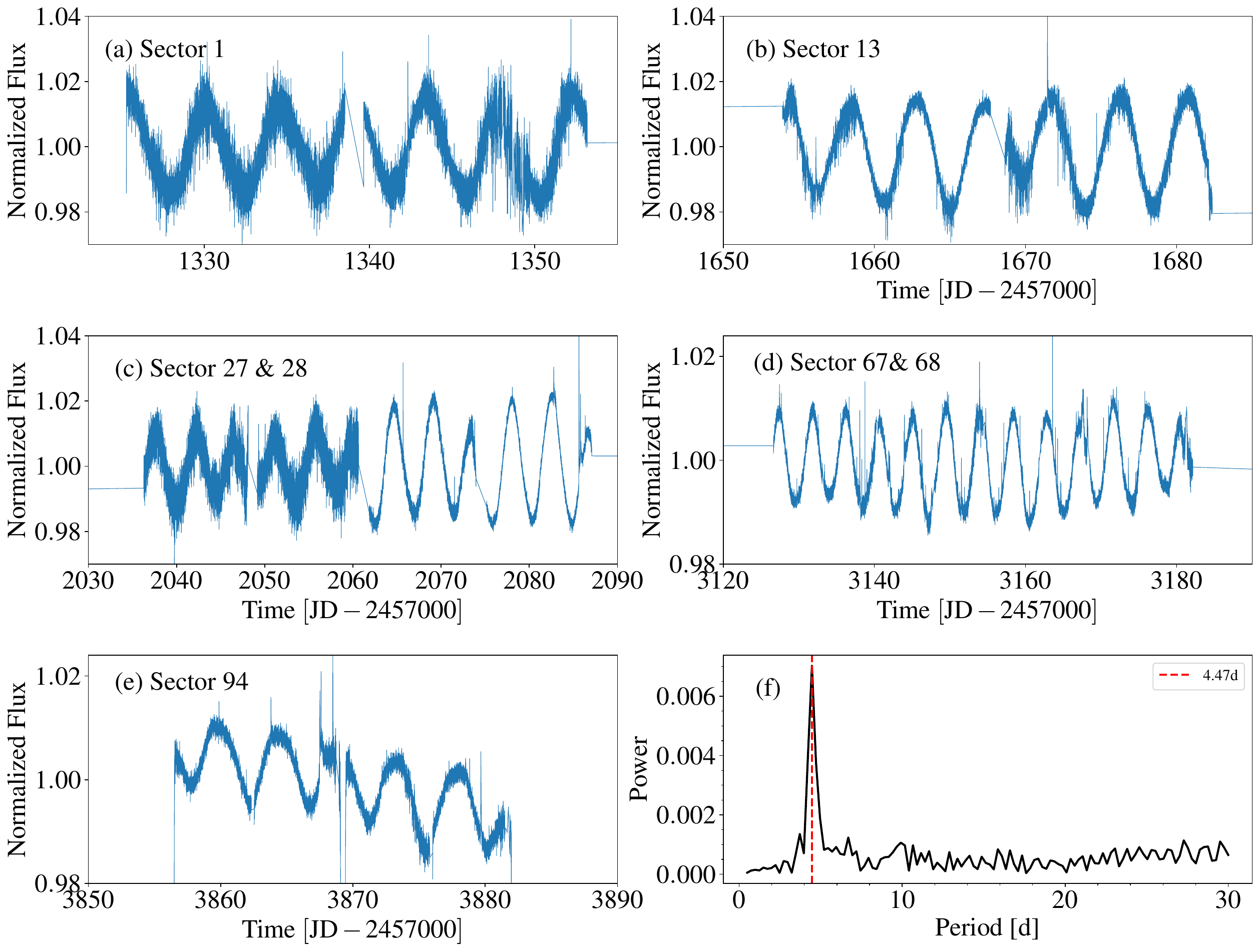}
    \caption{Panel a-e: TESS light curves of CPD-722713 obtained in Sector 01, 13,27,28,67,68 and 94. Panel f: the Lomb-Scargle periodogram of the detrended TESS lightcurves of CPD-722713. The highest peak in the periodogram is located at 4.47 days. }
    \label{fig:figurerot}
\end{figure}
\begin{figure*}
    \centering
    \includegraphics[width=0.8\linewidth]{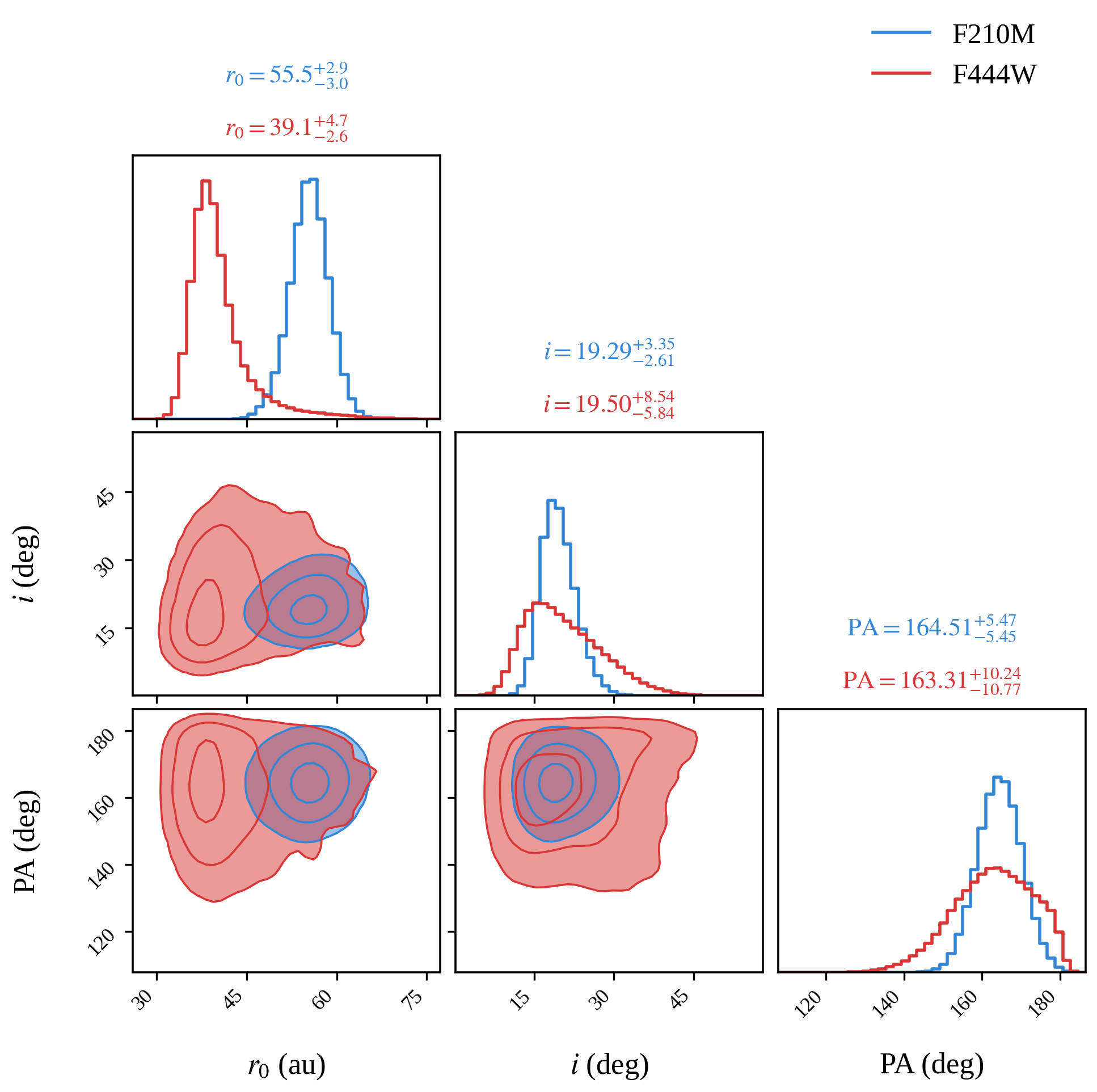}
    \caption{Posterior distributions of the disk radius $r_0$, inclination $i$, and position angle (PA) inferred from the F210M (blue) and F444W (red) observations. Contours correspond to $1$, $2$, and $3\sigma$ }
    \label{fig:corner_compare}
\end{figure*}

\newpage
\begin{figure*}
    \centering
    \includegraphics[width=\linewidth]{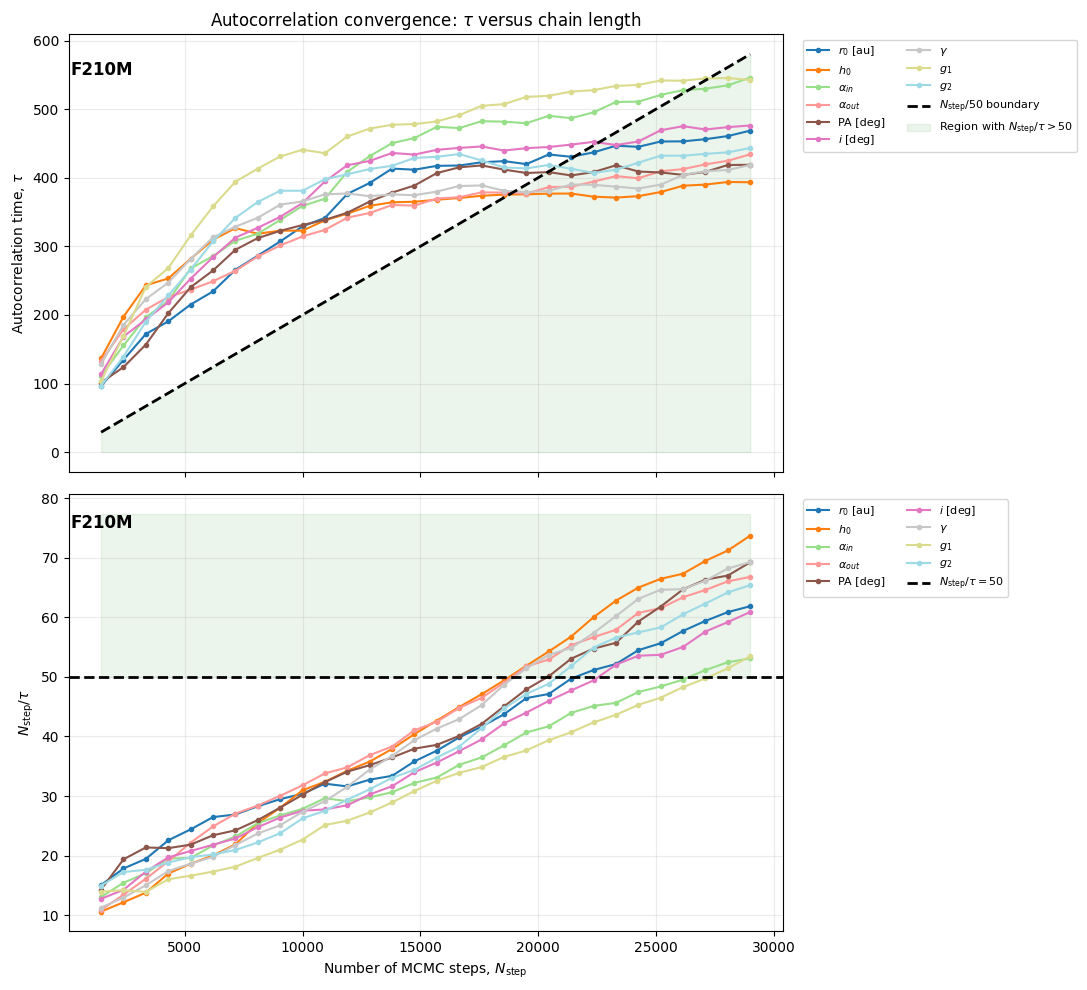}
    \caption{Autocorrelation time $\tau$ and ratio between total chain length and autocorrelation time $N_{\rm setp}/\tau$ as function of number of MCMC step $N_{\rm setp}$ for parameters in F210M filter. }
    \label{fig:tau210}
\end{figure*}

\newpage
\begin{figure*}
    \centering
    \includegraphics[width=1\linewidth]{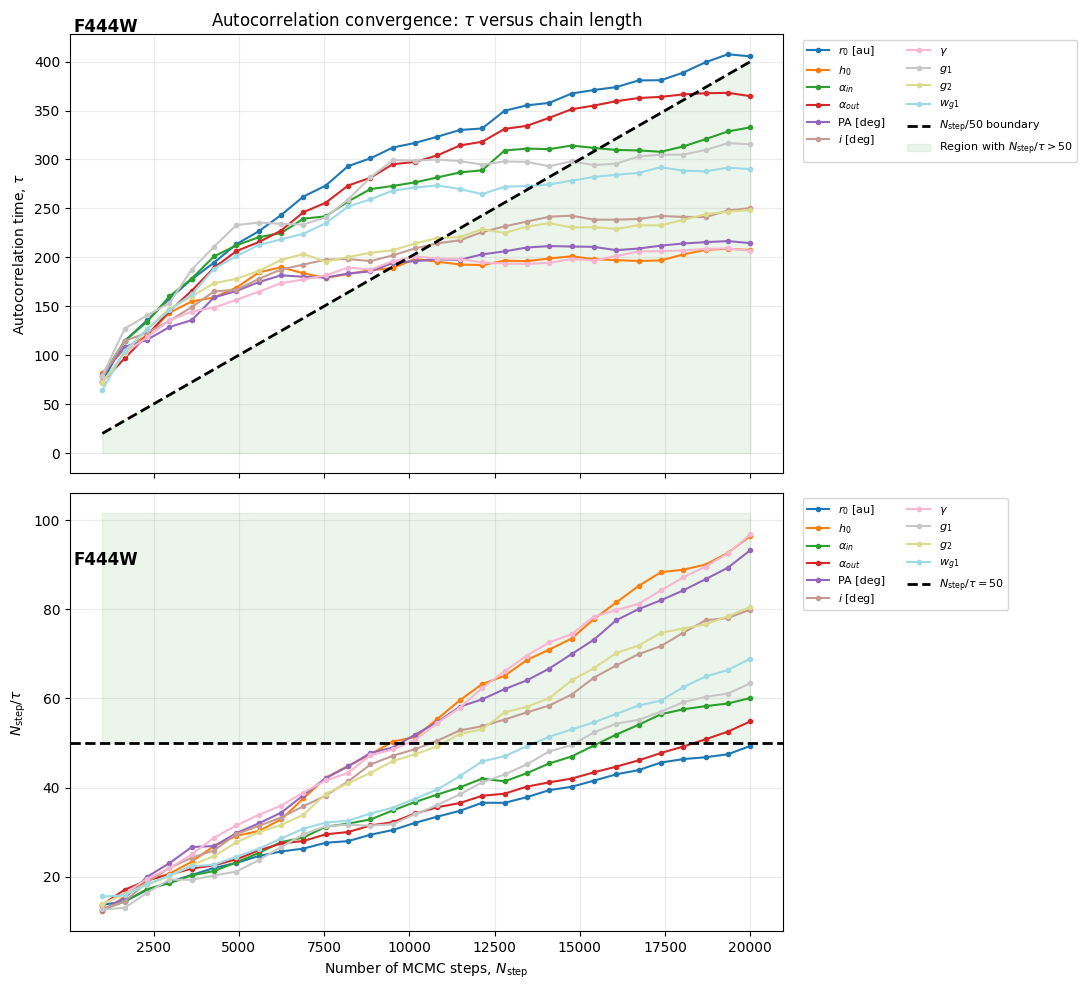}
    \caption{Autocorrelation time $\tau$ and ratio between total chain length and autocorrelation time $N_{\rm setp}/\tau$ as function of number of MCMC step $N_{\rm setp}$ for parameters in F444W filter.}
    \label{fig:tau444}
\end{figure*}
\newpage

\begin{figure*}
    \centering
    \includegraphics[width=1.1\linewidth]{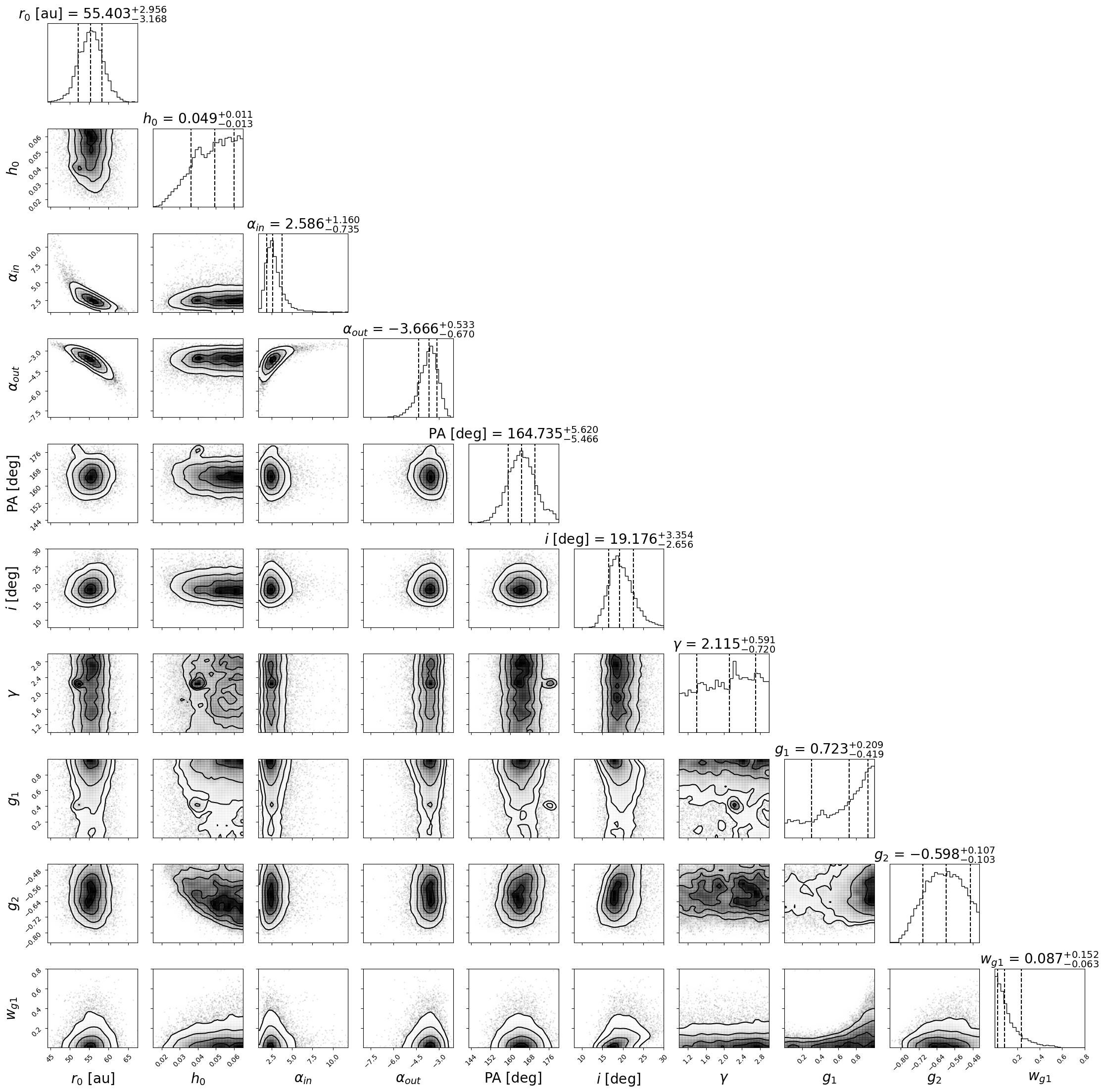}
    \caption{Joint posterior distributions for parameters used in the MCMC modeling of the disk observed in F210M filter. The values and histogram distributions of all parameters are shown, along with 1 $\sigma$ uncertainties.}
    \label{fig:figureA210}
\end{figure*}

\newpage
\begin{figure*}
    \centering
    \includegraphics[width=\linewidth]{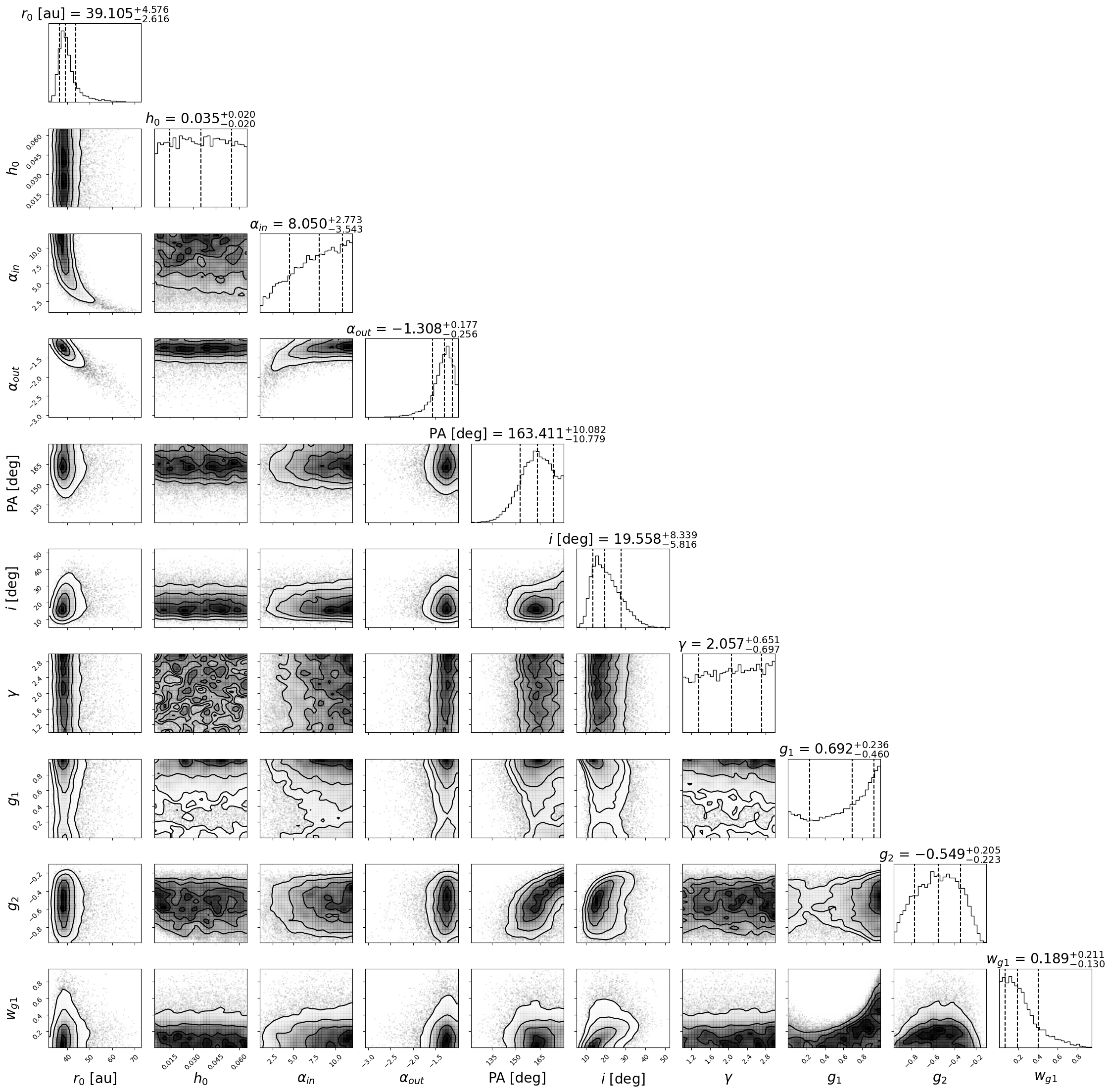}
    \caption{Joint posterior distributions for parameters used in the MCMC modeling of the disk observed in F444W filter. The values and histogram distributions of all parameters are shown, along with 1 $\sigma$ uncertainties.}
    \label{fig:figureA444}
\end{figure*}
\begin{figure}
    \centering
    \includegraphics[width=\linewidth]{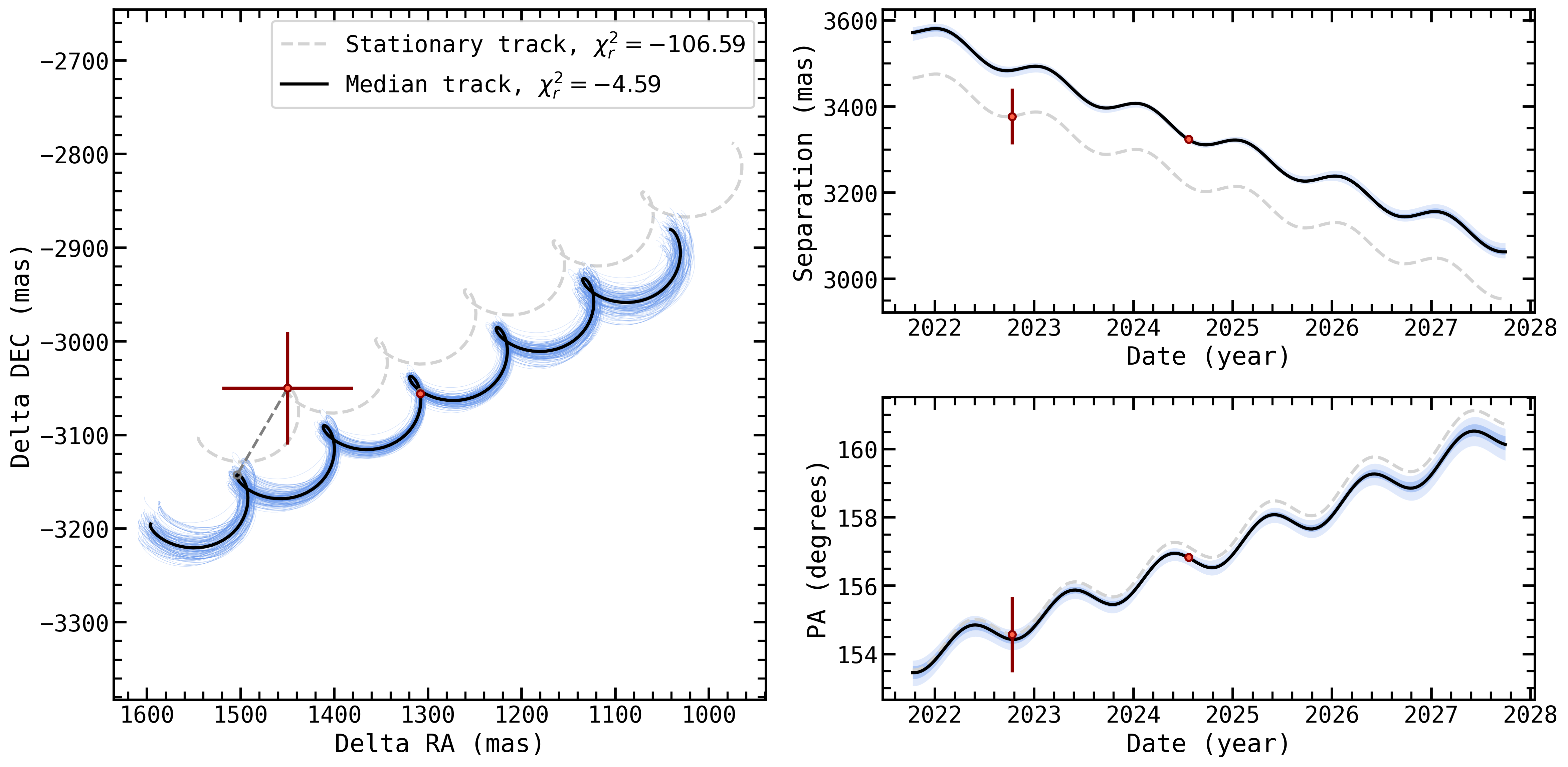}
    \caption{Common proper motion test with the astrometry with two epochs. Red points mark the two measurements. The gray dashed line shows the expected background track for a stationary source. Blue and black curves are MCMC fits that allow the background object to have its own motion.  }
    \label{fig:figureback}
\end{figure}

\newpage


\bibliography{sample63}{}
\bibliographystyle{aasjournal}


\end{CJK*}
\end{document}